\documentclass[amssymb,aps,prd,floatfix]{revtex4}
\usepackage[intlimits]{amsmath}
\usepackage{graphicx}
\usepackage{tensor}
\usepackage[dvipsnames]{xcolor}
\usepackage{hyperref}

\usepackage{dcolumn}
\usepackage{array}

\begin{document}

\author{Zuzanna Bakun}
\affiliation{Faculty of Physics, Astronomy and Applied Computer Science, Jagiellonian University, {\L}ojasiewicza 11, 30-348 Krak\'{o}w, Poland}

\author{Angelika {\L}ukanty}
\affiliation{Faculty of Physics, Astronomy and Applied Computer Science, Jagiellonian University, {\L}ojasiewicza 11, 30-348 Krak\'{o}w, Poland}

\author{Anastasiia Untilova}
\affiliation{Faculty of Physics, Astronomy and Applied Computer Science, Jagiellonian University, {\L}ojasiewicza 11, 30-348 Krak\'{o}w, Poland}

\author{Adam Cie\'{s}lik}
\affiliation{Institute of Theoretical Physics, Jagiellonian University, {\L}ojasiewicza 11, 30-348 Krak\'{o}w, Poland}

\author{Patryk Mach}
\affiliation{Institute of Theoretical Physics, Jagiellonian University, {\L}ojasiewicza 11, 30-348 Krak\'{o}w, Poland}

\title{Kerr Geodesics in horizon-penetrating Kerr coordinates: description in terms of Weierstrass functions}

\begin{abstract}
    We revisit the theory of timelike and null geodesics in the (extended) Kerr spacetime. This work is a sequel to a recent paper by Cie\'{s}lik, Hackmann, and Mach, who applied the so-called Biermann--Weierstrass formula to integrate Kerr geodesic equations expressed in Boyer--Lindquist coordinates. We show that a formulation based on the Biermann--Weierstrass theorem can also be applied in horizon-penetrating Kerr coordinates, resulting in solutions that are smooth across Kerr horizons. Horizon-penetrating Kerr coordinates allow for an explicit continuation of timelike and null geodesics between appropriate regions of the maximal analytic extension of the Kerr spacetime. A part of this work is devoted to a graphic visualisation of such geodesics.
\end{abstract}

\maketitle

\section{Introduction}

The Kerr metric is usually written in Boyer--Lindquist coordinates \cite{BoyerLindquist1967}. They lead to a simple form of the metric tensor with a single off-diagonal term---an advantage allowed by the fact that the Kerr metric is circular (or orthogonally transitive)---but suffer a coordinate singularity at the black hole horizon. A system of coordinates on the Kerr spacetime regular at the horizon has already been proposed in the original paper of Kerr in 1963 \cite{Kerr1963}. Here we will work with its more popular variant, obtained by performing an additional transformation $t = u - r$ (in the original notation of Kerr \cite{Kerr1963}) and a change in the convention regarding the black hole spin ($a \to -a$). These coordinates are often referred to simply as Kerr or Kerr--Schild coordinates \cite{rezzolla}. Since the latter term is usually reserved for a popular Cartesian-type coordinate system on the Kerr spacetime, we prefer to use the term ``horizon-penetrating Kerr coordinates.''

Separability of the geodesic motion in the Kerr spacetime was discovered in 1968 by Carter \cite{Carter1968} (see also \cite{carter_1968b,walker_penrose_1970}), who worked in the original coordinate system introduced by Kerr (save for the change $a \to - a$). Nowadays, the majority of works on Kerr geodesics use Boyer--Lindquist coordinates (a sample of papers published after 2000 includes \cite{schmidt_2002,glampedakis_kennefick_2002,Mino2003,teo_2003,drasco_hughes_2004,slezakova_2006,levin_perez_giz_2008,fujita_hikida_2009,levin_perez_giz_2009,perez_giz_levin_2009,hackmann_2010,hod_2013,grib_pavlov_vertogradov_2014,vertogradov_2015,lammerzahl_hackmann_2016,rana_mangalam_2019,tavlayan_tekin_2020,vandemeent_2020,stein_warburton_2020,gralla_lupsasca_2020,teo_2021,mummery_balbus_2022,mummery_balbus_2023,Dyson2023,gonzo_shi_2023}). While algebraic simplicity is an obvious advantage, using Boyer--Lindquist coordinates for Kerr geodesics can be, in some cases, misleading, as it leads to a non-physical near-horizon behavior---generic geodesics tend to wind up around the horizon. Boyer--Lindquist coordinates were also used in a recent paper \cite{CHM2023}, which introduced a uniform description of all generic timelike and null Kerr geodesics, based on Weierstrass functions. The advantage of this work was three-fold: Solutions were given in a form depending explicitly on the constants of motion and initial positions. No \textit{a priori} knowledge of radial turning points was needed. Finally, geodesics with no radial turning points were described by explicitly real formulas. 

In this paper we extend the analysis of \cite{CHM2023} to horizon-penetrating coordinates, which allow us to remove the singular behavior of geodesics at the black hole horizon, depending on the direction of motion. Quite surprisingly, little of the original simplicity of the analysis presented in \cite{CHM2023} and based on Boyer--Lindquist coordinates is lost. Radial and polar equations and solutions remain unchanged. Additional terms appear in the azimuthal and time equations, which can readily be integrated. All solutions can still be written in terms of standard Weierstrass functions, and a single set of formulas remains applicable to all generic geodesics. As in \cite{CHM2023}, solutions are fully specified by the values of constants of motion---the energy, the angular momentum, the Carter constant---and the initial position.

We show explicit examples of Kerr geodesics, attempting to visualize the behavior of those geodesics that cross the horizons. Such a visualization is especially tricky for geodesics which continue to negative values of the radius, allowed in the maximal analytic extension of the Kerr spacetime.

We prove that a future-directed timelike geodesic originating outside the event horizon and plunging into the black hole can, in horizon-penetrating Kerr coordinates, be continued smoothly through the event horizon. We also show that if this coordinate system admits a smooth transition of a given geodesic trajectory through one of horizons in a given radial direction, such a smooth transition is not allowed in the opposite direction.

\section{Geodesic equations in the Kerr spacetime}

\subsection{Metric conventions}

In this work, we employ standard geometric units in which the speed of light $c$ and the gravitational constant $G$ are set to $1$. The metric signature is $(-, +, +, +)$.

In Boyer--Lindquist coordinates $x^\mu = (t,r,\theta,\varphi)$ the Kerr metric can be written as
\begin{equation}
    g=-\left(1-\frac{2Mr}{\rho^{2}}\right) dt^{2} -\frac{4Mar\sin^{2}\theta}{\rho^{2}} dtd\varphi+ \frac{ \rho^2}{\Delta} dr^{2}+\rho^2 d\theta^{2}+\left(r^2+a^2+\frac{2Ma^2r \sin^2\theta}{\rho^2}\right) \sin^2\theta d\varphi^{2},
    \label{eqn:1}
\end{equation}
where
\begin{subequations}\label{eq:delta.ro}
\begin{eqnarray}
    \Delta & := & r^2 - 2Mr + a^2, \\
    \rho^2 & := & r^2 + a^2 \cos^2\theta.
\end{eqnarray}
\end{subequations}
We will only consider the case with $a^2 < M^2$, due to its physical relevance. Kerr horizons are located at the zeros of $\Delta$, i.e., at $r_\pm = M \pm \sqrt{M^2 - a^2}$. We will refer to horizons at $r = r_-$ and $r = r_+$ as the Cauchy and the event horizon, respectively.

The coordinate singularity at the horizons can be removed by a coordinate transformation to horizon-penetrating Kerr coordinates $(t^\prime,r,\theta,\varphi^\prime)$ defined in terms of one-forms
\begin{equation}
    dt^\prime = dt+\frac{2Mr}{\Delta}dr,   \quad
    d\varphi'=d\varphi+\frac{a}{\Delta}dr,
    \label{eqn:3}
\end{equation}
or more explicitly, by setting
\begin{equation}
    t'=t+2M\int\frac{rdr}{r^2-2Mr+a^2}, \qquad
    \varphi'=\varphi+a\int\frac{dr}{r^2-2Mr+a^2}.   
    \label{eqn:t.phi}
\end{equation}
In horizon-penetrating Kerr coordinates the Kerr metric reads
\begin{eqnarray}
    g & = & -\left(1-\frac{2Mr}{\rho^2}\right){dt^\prime}^2-\frac{4Mr}{\rho^2}a\sin^2\theta dt^\prime d\varphi^\prime + \frac{4Mr}{\rho^2} dt^\prime dr - 2 a \left(1 + \frac{2Mr}{\rho^2}\right)\sin^2\theta dr d\varphi^\prime
    \nonumber\\
    & & + \left(1+\frac{2Mr}{\rho^2}\right) dr^2 + \rho^2 d\theta^2 + \frac{\left[ (r^2+a^2)^2-a^2 \Delta \sin^2\theta \right]\sin^2\theta}{\rho^2} {d\varphi^\prime}^2.
    \label{gKerrSchild}
\end{eqnarray}

\begin{figure}
\begin{center}
\includegraphics[width=0.49
\textwidth]{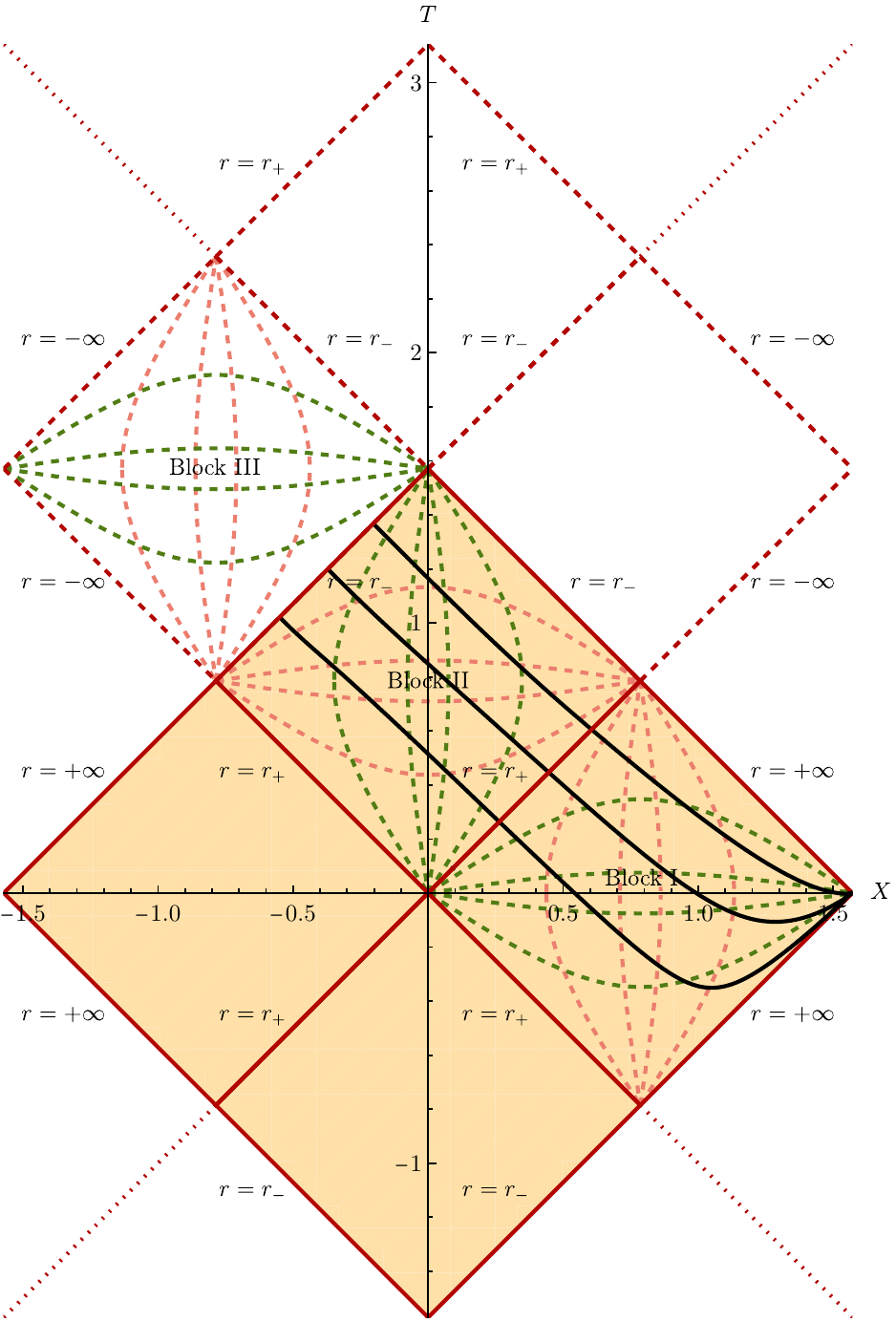}\includegraphics[width=0.49
\textwidth]{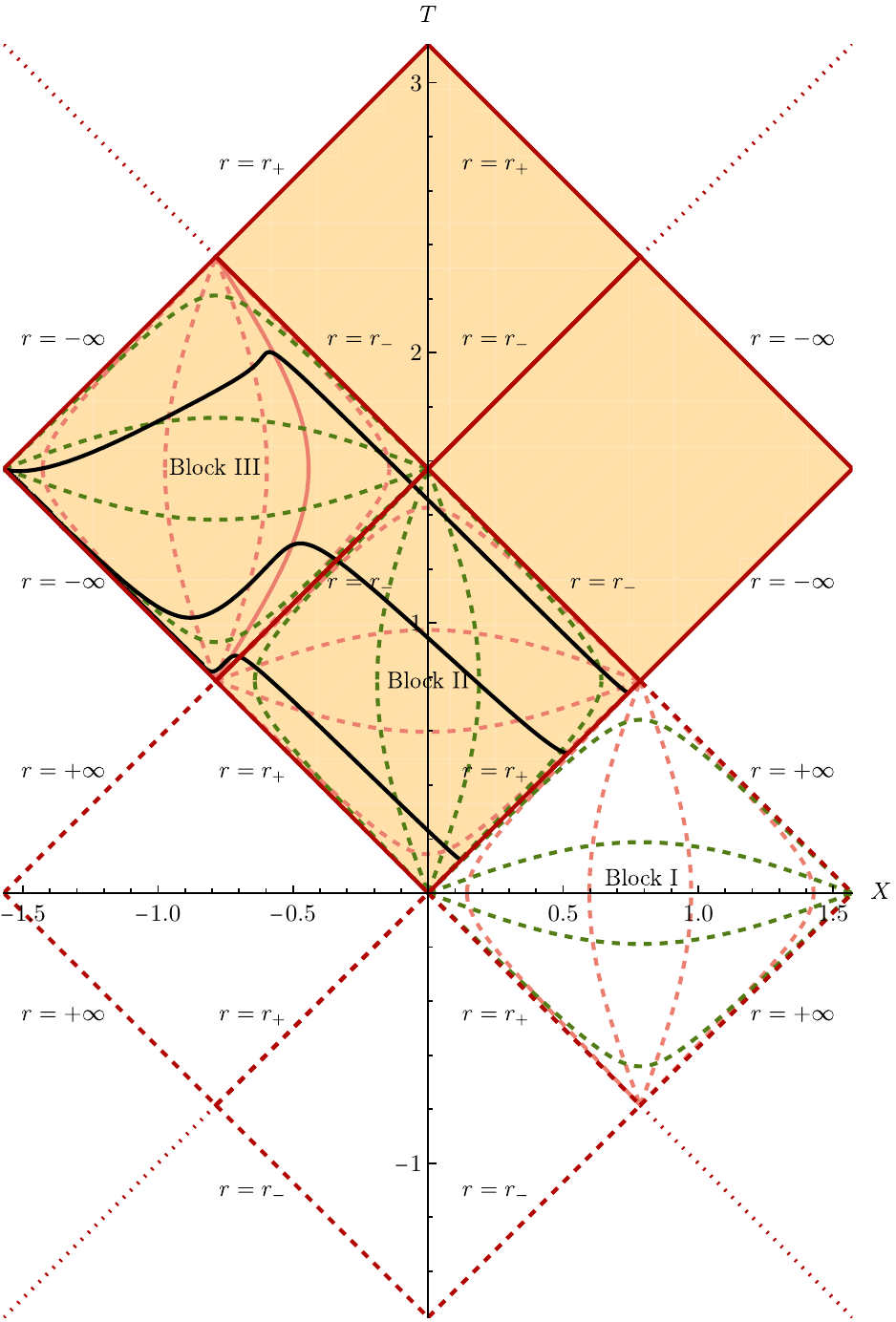}
\end{center}
\caption{\label{fig:penrose} Penrose diagrams of the symmetry axis of the Kerr spacetime for $a = 0.9 M$. Lines of constant time $t^\prime$ are plotted with solid black lines. We plot lines corresponding to (from the bottom to the top) $t^\prime = -5M, 0, 5M$. Surfaces of constant Boyer--Lindquist time $t$ are shown with green dashed lines. Orange dashed lines depict surfaces of constant tortoise radius $r_\ast$. A solid orange line within Block III depicts the surface with $r = 0$. Left and right panels correspond, respectively, to coordinate systems $K$ and $K'$, described in Appendix \ref{appendix:penrose}. Block II can be represented in both coordinate systems, but the graphs of lines of constant $t'$ depend on the chosen coordinate system, and consequently they are different in both panels.}
\end{figure}

In this paper we will restrict ourselves to a region of the maximally extended Kerr spacetime covered by horizon-penetrating Kerr coordinates with the ranges $t^\prime \in \mathbb R$, $r \in \mathbb R$, $\theta \in [0,\pi]$, $\varphi \in [0,2\pi)$. It spans across three causal diamonds or Boyer--Lindquist blocks, numbered as I, II, and III, according to a convention used in \cite{Neill1995}. They are defined by the following ranges of the radius. Block I: $r > r_+$; Block II: $r_- < r < r_+$; Block III: $r < r_-$. We depict this region on two Penrose conformal diagrams corresponding to the symmetry axis of the Kerr spacetime in Fig.\ \ref{fig:penrose}. To show the properties of the time foliation defined by the horizon-penetrating Kerr coordinates, we plot, in Fig.\ \ref{fig:penrose}, hypersurfaces of constant time $t^\prime$. They remain regular across the event horizon $r = r_+$ joining Blocks I and II and across the Cauchy horizon $r = r_-$, joining Blocks II and III. Since the standard Kruskal construction of conformal diagrams does not allow for a single regular coordinate system covering all three Blocks I, II, and III, we show in Fig.\ \ref{fig:penrose} two separate diagrams: a diagram with a regular coordinate patch covering Blocks I and II (left panel) and another diagram with a regular coordinate patch covering Blocks II and III (right panel). Further details of the construction of these diagrams are given in Appendix \ref{appendix:penrose}.

\subsection{Geodesic equations}

Geodesic equations can be written in the Hamiltonian form as
\begin{equation}
\label{eq:ham}
\frac{dx^\mu}{{d\tilde\tau}}=\frac{\partial H}{\partial p_\mu}, \quad \frac{dp_\nu}{{d\tilde\tau}}=-\frac{\partial H}{\partial x^\nu},
\end{equation}
where $p^\mu=dx^\mu/d\tilde\tau$, $H(x^\alpha,p_\beta)=\frac{1}{2}g^{\mu\nu}(x^\alpha)p_\mu p_\nu=-\frac{1}{2}m^2$, and $m$ is the particle rest mass. The four-velocity $u^\mu = d x^\mu/d\tau$ is normalised as $g_{\mu\nu}u^\mu u^\nu=-\delta_1$, where
\begin{equation}
    \delta_1 = \begin{cases} 1 & \text{for timelike geodesics}, \\0 & \text{for null geodesics}. \end{cases}
\end{equation}
The above conventions imply that the affine parameter $\tilde \tau$ and the proper time $\tau$ are related by $\tilde \tau = \tau/m$.

Let the metric $g$ be given by Eq.\ (\ref{eqn:1}) or (\ref{gKerrSchild}). A standard reasoning shows that $H$, $E := - p_t$ and $l_z := p_\varphi$ are constants of motion. The fourth constant, referred to as the Carter constant $\mathcal K$, can be derived by a separation of variables in the corresponding Hamilton--Jacobi equation \cite{Carter1968}. This separation can be performed both in Boyer--Lindquist and in horizon-penetrating Kerr coordinates. Moreover, in both coordinate systems constants $H$, $E$, $l_z$, and $\mathcal K$ have the same numerical values. In particular, momentum components $p_t$ and $p_\varphi$ transform as $p_{t^\prime} = p_t = -E$ and $p_{\varphi^\prime} = p_\varphi = l_z$.

In Boyer--Lindquist coordinates geodesic equations can be written as
\begin{subequations}
\label{eq:BLeqs}
\begin{eqnarray}
\label{eq:BLradial}
    \rho^2\frac{dr}{{d\tilde\tau}} & = & \epsilon_r\sqrt{R(r)},\\
    \rho^2\frac{d\theta}{{d\tilde\tau}} & = & \epsilon_\theta\sqrt{\Theta(\theta)},\\
    \rho^2\frac{d\varphi}{{d\tilde\tau}} & = & \frac{a[(r^2+a^2)E-al_z]}{\Delta}+\frac{1}{\sin^2\theta}(l_z-aE\sin^2\theta),\\
    \rho^2\frac{dt}{{d\tilde\tau}} & = & \frac{(r^2+a^2)[(r^2+a^2)E-al_z]}{\Delta}+a(l_z-aE\sin^2\theta),
\end{eqnarray}
\end{subequations}
where we denoted
\begin{subequations}
\begin{eqnarray}
    R(r) & := & [(r^2+a^2)E-al_z]^2-\Delta(m^2r^2+\mathcal K), \\
    \Theta(\theta) & := & \mathcal K-m^2a^2\cos^2\theta- \left( \frac{l_z}{\sin\theta}-a\sin\theta E \right)^2.
\end{eqnarray}
\end{subequations}
We will refer to $R(r)$ and $\Theta(\theta)$ as the radial and polar effective potentials, respectively. The signs $\epsilon_r = \pm 1$ and $\epsilon_\theta = \pm 1$ indicate the direction of motion with respect to the radial and polar coordinates. Strictly speaking, Eqs.\ (\ref{eq:BLeqs}) constitute a set of 4 first-order equations, while there are 8 first-order geodesic equations (\ref{eq:ham}). There exist solutions of Eqs.\ (\ref{eq:BLeqs}), sometimes referred to as ``singular'', that do not correspond to geodesics and do not satisfy Eqs.\ (\ref{eq:ham}). For instance, if $r_0$ is a multiplicity one zero of $R(r)$, then $r \equiv r_0$ would satisfy Eq.\ (\ref{eq:BLradial}), but it would not correspond to a geodesic.

There is a useful parametrization of Kerr geodesics, introduced by Mino in \cite{Mino2003}, which allows to partially decouple Eqs.\ (\ref{eq:BLeqs}). The so-called Mino time $\tilde s$ is defined by
\begin{equation}
    \rho^2\frac{dx^\mu}{{d\tilde\tau}}=\frac{dx^\mu}{{d\tilde s}}
\end{equation}
or
\begin{equation}
    \tilde\tau = \int\limits_0^{\tilde{s}} \rho^{2}ds.
\end{equation}
Using $\tilde s$ as the geodesic parameter, we obtain
\begin{subequations}
\label{eq:BLmino}
\begin{eqnarray}
    \frac{dr}{{d\tilde s}} & = & \epsilon_r\sqrt{R(r)}, \\
    \frac{d\theta}{{d\tilde s}} & = & \epsilon_\theta\sqrt{\Theta(\theta)}, \\
    \frac{d\varphi}{{d\tilde s}} & = & \frac{a[(r^2+a^2)E-al_z]}{\Delta}+\frac{1}{\sin^2\theta}(l_z-aE\sin^2\theta), \\
    \frac{dt}{{d\tilde s}} & = & \frac{(r^2+a^2)[(r^2+a^2)E-al_z]}{\Delta}+a(l_z-aE\sin^2\theta).
\end{eqnarray}
\end{subequations}

Geodesic equations in the horizon-penetrating Kerr coordinates can be obtained simply by the following vector transformation
\begin{subequations}
\begin{eqnarray}
    \frac{dt'}{{d\tilde s}} & = & \frac{\partial t'}{\partial t}\frac{dt}{d \tilde s}+\frac{\partial t'}{\partial r}\frac{dr}{d \tilde s}=\frac{dt}{d\tilde s}+\frac{2Mr}{\Delta}\frac{dr}{d\tilde s}, \\
    \frac{d\varphi'}{{d\tilde s}} & = & \frac{\partial \varphi'}{\partial \varphi}\frac{d \varphi}{d \tilde s}+\frac{\partial \varphi '}{\partial r}\frac{dr}{d \tilde s}=\frac{d\varphi}{d\tilde s}+\frac{a}{\Delta}\frac{dr}{d\tilde s}.
\end{eqnarray}
\end{subequations}
This gives
\begin{subequations}
\label{eq:KSeqs}
\begin{eqnarray}
    \frac{dr}{{d\tilde s}} & = & \epsilon_r\sqrt{R(r)},\\
    \frac{d\theta}{{d\tilde s}} & = & \epsilon_\theta\sqrt{\Theta(\theta)},\\
    \frac{d\varphi'}{{d\tilde s}} & = & \frac{a[(r^2+a^2)E-al_z]}{\Delta}+\frac{1}{\sin^2\theta}(l_z-aE\sin^2\theta)+\frac{a}{\Delta}\epsilon_r\sqrt{R(r)},\\
    \frac{dt'}{{d\tilde s}} & = & \frac{(r^2+a^2)[(r^2+a^2)E-al_z]}{\Delta}+a(l_z-aE\sin^2\theta)+\frac{2Mr}{\Delta}\epsilon_r\sqrt{R(r)}.
\end{eqnarray}
\end{subequations}

In the remainder of this paper we will use the following dimensionless variables:
\begin{equation}
\label{dimensionless}
    a=M\alpha, \qquad t'=MT',\quad r=M\xi,\quad E=m\varepsilon,\quad \mathcal K=M^2m^2\kappa, \quad l_z=Mm\lambda_z, \quad \tilde s=\frac{1}{Mm}s.
\end{equation}
Dimensionless radii corresponding to Kerr horizons will be denoted by $\xi_\pm = 1 \pm \sqrt{1 - \alpha^2}$. For null geodesics, for which $m = 0$, the parameter $m$ in the above equations can be replaced with any mass parameter $\tilde m > 0$.

In dimensionless variables (\ref{dimensionless}), geodesic equations (\ref{eq:KSeqs}) have the form
\begin{subequations}\label{eq:mot}
\begin{eqnarray}\label{eq:mot.ksi}
    \frac{d\xi}{{ds}} & = &\epsilon_r\sqrt{\tilde R}, \\ \label{eq:mot.th}
    \frac{d\theta}{{ds}} & = &\epsilon_\theta\sqrt{\tilde\Theta}, \\ \label{eq:motphi}
    \frac{d\varphi'}{{ds}} & = & \frac{\alpha[(\xi^2+\alpha^2)\varepsilon-\alpha \lambda_z]}{\xi^2-2\xi+\alpha^2}+\frac{1}{\sin^2\theta}(\lambda_z-\alpha\varepsilon\sin^2\theta)+\frac{\alpha}{\xi^2-2\xi+\alpha^2}\epsilon_r\sqrt{\tilde R}, \\ \label{eq:mott}
    \frac{dT'}{{ds}} & = & \frac{(\xi^2+\alpha^2)[(\xi^2+\alpha^2)\varepsilon-\alpha\lambda_z]}{\xi^2-2\xi+\alpha^2}+\alpha(\lambda_z-\alpha\varepsilon\sin^2\theta)+\frac{2\xi}{\xi^2-2\xi+\alpha^2}\epsilon_r\sqrt{\tilde R},
\end{eqnarray}
\end{subequations}
where
\begin{subequations}
\begin{eqnarray}
    \tilde R & := & [(\xi^2+\alpha^2)\varepsilon-\alpha\lambda_z]^2-(\xi^2-2\xi+\alpha^2)(\delta_1\xi^2+\kappa),
\label{eq:R} \\
    \tilde\Theta & := & \kappa-\delta_1\alpha^2\cos^2\theta-\frac{1}{\sin^2\theta}(\lambda_z-\alpha\varepsilon\sin^2\theta)^2.
\end{eqnarray}
\end{subequations}
Note that the coordinate transformation from Boyer--Lindquist to horizon-penetrating Kerr coordinates affects only the equations for $\varphi^\prime$ and $T^\prime$, while the radial and polar equations remain unchanged. Equations (\ref{eq:mot}) can also be obtained directly by working in the horizon-penetrating Kerr coordinates and by separating the variables in the Hamilton--Jacobi equation. We emphasise a connection with the Boyer--Lindquist form (\ref{eq:BLmino}), in order to make use of solutions to Eqs.\ (\ref{eq:BLmino}) derived in \cite{CHM2023}. In the next section, we review the solutions of the radial and polar equations obtained in \cite{CHM2023} and focus on equations for $\varphi^\prime$ and $T^\prime$.

\section{Solutions of geodesic equations}

\subsection{Biermann--Weierstrass formula}

The form of solutions used in this work predominantly rely on the following result due to Biermann and Weierstrass \cite{biermann_1865}. Proofs of this theorem can be found in \cite{Greenhill_1892, Reynolds_1989, CM2022}.

Let $f$ be a quartic polynomial
\begin{equation}
 f(x) = a_0 x^4 + 4 a_1 x^3 +6 a_2 x^2 + 4a_3 x + a_4,
\end{equation}
and let $g_2$ and $g_3$ denote Weierstrass invariants of $f$:
\begin{subequations}
\label{invariants_theorem}
\begin{eqnarray}
        g_2 & = &  a_0 a_4 - 4a_1 a_3 + 3 a_2^2, \\
        g_3 & = & a_0 a_2 a_4 + 2a_1 a_2 a_3 -a_2^3 -a_0 a_3^2 - a_1^2 a_4.
\end{eqnarray}
\end{subequations}
Denote
\begin{equation}
\label{zthm}
 z(x) = \int^x_{x_0} \frac{dx^\prime}{\sqrt{f(x^\prime)}},
\end{equation}
where $x_0$ can be any constant. Then $x$ can be expressed as
\begin{equation}
\label{glowne}
x = x_0 + \frac{- \sqrt{f(x_0)} \wp'(z)
  + \frac{1}{2} f'(x_0) \left[ \wp(z) - \frac{1}{24}f''(x_0) \right]
  + \frac{1}{24} f(x_0) f'''(x_0)  }{2 \left[ \wp(z)
  - \frac{1}{24} f''(x_0) \right]^2 - \frac{1}{48} f(x_0) f^{(4)}(x_0) },
\end{equation}
where $\wp(z) =\wp(z;g_2,g_3)$ is the Weierstrass function with invariants
(\ref{invariants_theorem}). In addition
\begin{subequations}
\begin{eqnarray}
    \wp(x) & = &\frac{\sqrt{f(x)f(x_0)}+f(x_0)}{2(x-x_0)^2}+\frac{f'(x_0)}{4(x-x_0)}+\frac{f''(x_0)}{24}, \\
    \wp' (x) & = & -\left[\frac{f(x)}{(x-x_0)^3}-\frac{f'(x_0)}{4(x-x_0)}\right]\sqrt{f(x_0)} - \left[\frac{f(x)}{(x-x_0)^3}+\frac{f'(x_0)}{4(x-x_0)} \right]\sqrt{f(x)}.
\end{eqnarray} 
\end{subequations}

\subsection{Radial motion}

The dimensionless radial potential $\tilde R$ can be written as
\begin{equation}
    \tilde R(\xi) = a_0 \xi^4 + 4 a_1 \xi^3 + 6 a_2 \xi^2 + 4a_3 \xi + a_4,
\end{equation}
where
\begin{subequations}
\label{radialcoeffs}
\begin{eqnarray}
        a_0 & = & \varepsilon^2 - \delta_1, \\
        a_1 & = & \frac{1}{2} \delta_1, \\
        a_2 & = & -\frac{1}{6}(\delta_1 \alpha^2 + \kappa - 2\alpha^2 \varepsilon^2 + 2 \alpha \varepsilon \lambda_z) , \\
        a_3 & = & \frac{1}{2} \kappa, \\
        a_4 & = & \alpha^4 \varepsilon^2 - \alpha^2 \kappa - 2 \alpha^3 \varepsilon \lambda_z + \alpha^2 \lambda_z^2 = -\alpha^2 [\kappa - (\alpha \varepsilon - \lambda_z)^2].
\end{eqnarray}
\end{subequations}
Weierstrass invariants associated with the coefficients (\ref{radialcoeffs}) will be denoted by
\begin{subequations}
\label{radialinvariants}
\begin{eqnarray}
    g_{\tilde R,2} & = & a_0 a_4 - 4 a_1 a_3 + 3 a_2^2, \\
    g_{\tilde R,3} & = & a_0 a_2 a_4 + 2 a_1 a_2 a_3 - a_2^3 - a_0 a_3^2 - a_1^2 a_4.
\end{eqnarray}
\end{subequations}
A direct application of the Biermann--Weierstrass theorem to Eq.\ (\ref{eq:mot.ksi}) yields the formula for $\xi=\xi(s)$,
\begin{equation}
\label{xisol}
    \xi(s)=\xi_0+\frac{-\epsilon_{r,0}\sqrt{\tilde R(\xi_0)} \wp'_{\tilde R}(s)+\frac{1}{2}\tilde R'\xi_0)[\wp_{\tilde R}(s)-\frac{1}{24}\tilde R''(\xi_0)]+\frac{1}{24}\tilde R(\xi_0)\tilde R'''(\xi_0)]}{2[\wp_{\tilde R}(s)-\frac{1}{24}\tilde R''(\xi_0)]^2-\frac{1}{48}\tilde R(\xi_0) \tilde R^{(4)}(\xi_0)},
\end{equation}
where $\wp_{\tilde R}(s)=\wp_{\tilde R}(s;g_{\tilde R,2}g_{\tilde R,3})$ is the Weierstrass function with invariants (\ref{radialinvariants}), and $\xi_0 = \xi(0)$ is an arbitrarily selected initial radius corresponding to $s = 0$. In Equation (\ref{xisol}) the sign $\epsilon_{r,0} = \pm 1$ denotes a value of $\epsilon_r$ corresponding to the initial location $\xi_0$. In other words, $\epsilon_{r,0}$ is a part of initial data (initial parameter), while the sign $\epsilon_r$ can change along a given geodesic.

\subsection{Polar motion}

Equation (\ref{eq:mot.th}) can be transformed to the Biermann--Weierstrass form by a substitution $\mu=\cos \theta$, which provides a one to one mapping for $0 \leq \theta \leq \pi$. Defining $g(\mu) := \sin^2 \theta \tilde\Theta(\theta)$, we get
\begin{equation}
    \frac{d\mu(s)}{ds} = - \epsilon_\theta \sqrt{g(\mu)}.
\end{equation}
The function $g(\mu)$ is a polynomial with respect to $\mu$ given by
\begin{equation}
    g(\mu) = b_0 \mu^4 + 6 b_2 \mu^2 + b_4,
\end{equation}
where the coefficients $b_0$, $b_2$, and $b_4$ can be expressed as
\begin{subequations}
\begin{eqnarray}
    b_0 & = & - \alpha^2 \left( \varepsilon^2 - \delta_1 \right) = - \alpha^2 a_0, \\
    b_2 & = & \frac{1}{6} \left( - \alpha^2 \delta_1 + 2 \alpha^2 \varepsilon^2 - \kappa - 2 \alpha \varepsilon \lambda_z \right) = a_2, \\
    b_4 & = & - \alpha^2 \varepsilon^2 + \kappa + 2 \alpha \varepsilon \lambda_z - \lambda_z^2 = \kappa - (\alpha \varepsilon - \lambda_z)^2 = - \frac{a_4}{\alpha^2},
\end{eqnarray}
\end{subequations}
and $a_0$, $a_2$, and $a_4$ are given by Eqs.\ (\ref{radialcoeffs}). Weierstrass invariants associated with coefficients $b_0$, $b_2$, and $b_4$ can be written as
\begin{subequations}
\begin{eqnarray}
g_{g,2} & = & b_0 b_4 + 3 b_2^2, \\
g_{g,3} & = & b_0 b_2 b_4 - b_2^3.
\end{eqnarray}
\end{subequations}

Again, using the Biermann--Weierstrass formula, one can write the expression for $\mu=\mu(s)$ in the form
\begin{equation}
\label{musol}
\mu(s)=\mu_0+\frac{\epsilon_{\theta ,0}\sqrt{g(\mu_0)} \wp'_{g}(s)+\frac{1}{2}g'(\mu_0)[\wp_{g}(s)-\frac{1}{24}g''(\mu_0)]+\frac{1}{24}g(\mu_0)g'''(\mu_0)]}{2[\wp_{g}(s)-\frac{1}{24}g''(\mu_0)]^2-\frac{1}{48}g(\mu_0) g^{(4)}(\mu_0)},
\end{equation}
where $\wp_{g}(s)=\wp_{g}(s;g_{g,2}g_{ g,3})$ and $\mu_0 = \mu(0) = \cos \theta_0 =\cos \theta(0)$ represents an initial value corresponding to $s=0$. As in Eq.\ (\ref{xisol}), the sign $\epsilon_{\theta ,0} = \pm 1$ is a parameter equal to $\epsilon_\theta$ at $\theta = \theta_0$, and it remains constant along the entire geodesic. A more detailed discussion of solutions (\ref{xisol}) and (\ref{musol}) can be found in \cite{CHM2023}.

\subsection{Azimuthal motion}

Equation (\ref{eq:motphi}) consists of a component related to the azimuthal motion in Boyer--Lindquist coordinates and an additional term, dependent on $\xi(s)$, arising from transformation (\ref{eqn:t.phi}). We will first treat the two components separately and then discuss their sum, which may remain regular across the horizons. An integration with respect to the Mino time $s$ yields
\begin{eqnarray}
    \varphi^\prime(s)-\varphi^\prime(0) & = & \int_{0}^{s} \left\{ \frac{\alpha [(\xi^2(\bar{s})+\alpha^2)\varepsilon-\alpha \lambda_z]}{\xi^2(\bar{s})-2\xi(\bar{s})+\alpha^2}+\frac{1}{\sin^2\theta(\bar{s})} \left[ \lambda_z-\alpha\varepsilon\sin^2\theta(\bar{s})\right] \right\} d\bar{s}+\int_{0}^{s} \frac{ \alpha \epsilon_r \sqrt{\tilde R(\bar{s})}}{\xi^2(\bar{s})-2\xi(\bar{s})+\alpha^2} d\bar{s} \label{eq:phi1} \\
    & = & \alpha \int_{0}^{s} \frac{2 \xi(\bar s) \varepsilon - \alpha \lambda_z}{\xi^2(\bar{s})-2\xi(\bar{s})+\alpha^2} d\bar{s} + \lambda_z \int_0^s \frac{d \bar s}{\sin^2 \theta(\bar s)} + \alpha \int_{0}^{s} \frac{ \epsilon_r \sqrt{\tilde R(\bar{s})}}{\xi^2(\bar{s})-2\xi(\bar{s})+\alpha^2} d\bar{s} \nonumber \\
    & = & J_\mathrm{BL}(s) + J_\xi(s),
\end{eqnarray}
where $J_\mathrm{BL}(s) := \tilde J_\mathrm{BL}(s) + J_\theta(s)$ and
\begin{eqnarray}
    \tilde J_\mathrm{BL}(s) & := & \alpha \int_{0}^{s} \frac{2 \xi(\bar s) \varepsilon - \alpha \lambda_z}{\xi^2(\bar{s})-2\xi(\bar{s})+\alpha^2} d\bar{s}, \label{eqn:BL} \\
    J_\theta(s) & := & \lambda_z \int_0^s \frac{d \bar s}{\sin^2 \theta(\bar s)}, \\
    J_{\xi}(s) & := & \alpha \int_{0}^{s} \frac{\epsilon_r  \sqrt{\tilde R (\bar{s})}}{\xi^2(\bar{s})-2\xi(\bar{s})+\alpha^2} d\bar{s}. \label{eqn:Jks}
\end{eqnarray}

Integrals $\tilde J_\mathrm{BL}$ and $J_\theta$ were expressed in terms of Weierstrass functions in \cite{CHM2023}. To obtain a solution for Eq.\ (\ref{eqn:Jks}), we write
\begin{equation}
    J_{\xi}(s)= \alpha \int_{0}^{s} \frac{ d \xi(\bar{s})/d \bar s}{\xi^2(\bar{s})-2\xi(\bar{s})+\alpha^2} d\bar{s}.
\label{eqn:Jksi}
\end{equation}
For a segment of a geodesic along which $\xi = \xi(s)$ is monotonic, one change the integration variable to $\xi$ and write
\begin{equation}
\label{Jxisol}
    J_{\xi}(s)= \alpha \int_{\xi_0}^{\xi(s)} \frac{d\bar{\xi}}{\bar{\xi}^2-2 \bar{\xi}+\alpha^2}=\alpha \left[ \frac{ \arctan \left(\frac{\xi(s)-1}{\sqrt{\alpha^2 -1}}\right)}{\sqrt{\alpha^2 -1}}- \frac{ \arctan \left(\frac{\xi_0 - 1}{\sqrt{\alpha^2 -1}}\right)}{\sqrt{\alpha^2 -1}} \right],
\end{equation}
which, together with Eq.\ (\ref{xisol}), provides the solution. Note that due to the symmetry of Eq.\ (\ref{eq:mot.ksi}), expression (\ref{Jxisol}) remains valid also for trajectories passing through radial turning points.

The integrals $J_\mathrm{BL}$ and $J_\xi$ are, generically, divergent at the horizons, i.e., for $s$ such that $\xi(s) = \xi_\pm$, but the sum $J_\mathrm{BL} + J_\xi$ can remain regular. A direct calculation shows that $J_\text{BL}(s)$ and $J_\xi (s)$ can be expressed in the form
\begin{equation}
\label{regularJ}
    \tilde J_\mathrm{BL}(s)+J_\xi (s) = \alpha \int_{0}^{s} \left\{ \frac{ \delta_1 \xi ^2(\bar{s})+\kappa}{[\xi^2(\bar{s})+\alpha^2]\varepsilon -\alpha \lambda_z - d \xi(\bar s)/d \bar{s}} - \varepsilon \right \}d\bar{s}.
\end{equation}
The integrand in Eq.\ (\ref{regularJ}) can be divergent, if
\begin{equation}
[\xi^2(s)+\alpha^2] \varepsilon -\alpha \lambda_z - \frac{d \xi(s)}{d s} = 0
\end{equation}
or, equivalently,
\begin{equation}[\xi^2(s)+\alpha^2]\varepsilon -\alpha \lambda_z = \epsilon_r \sqrt{\tilde R}.
\end{equation}
By computing the square of the above equation we see that this can only happen, if
\[ [\xi^2(s) - 2 \xi(s) + \alpha^2][\delta_1 \xi^2(s) + \kappa] = 0, \]
that is at $\xi(s) = \xi_\pm$. On the other hand, the expression $[\xi^2(s)+\alpha^2]\varepsilon -\alpha \lambda_z - d \xi(s)/d s$ can be clearly non-zero at the horizon, depending on the sign $\epsilon_r$. In our examples discussed in Sec.\ \ref{sec:examples} this happens for $\epsilon_r = -1$, i.e., for incoming geodesics. In general, if the sign of $[\xi^2(s) + \alpha^2] \varepsilon - \alpha \lambda_z$ can be controlled, one can exclude the possibility that $\tilde J_\mathrm{BL}$ diverges at the horizon. For instance, for  $[\xi^2(s) + \alpha^2] \varepsilon - \alpha \lambda_z > 0$ and $\epsilon_r = -1$, the denominator in Eq.\ (\ref{regularJ}) remains strictly positive. We discuss this problem in more detail in Sec.\ \ref{sec:regularity}.

\subsection{Time coordinate}

Similarly to the equation for $\varphi^\prime$, the right-hand side of Eq.\ (\ref{eq:mott}) also consist of a Boyer--Lindquist term and a term associated with the transformation to horizon-penetrating Kerr coordinates. Integrating Eq.\ (\ref{eq:mott}) one gets
\begin{eqnarray}
    T^\prime(s)-T^\prime(0) & = & \int_{0}^{s} \left\{ \frac{[\xi^2(\bar{s})+\alpha^2][(\xi^2(\bar{s})+\alpha^2)\varepsilon-\alpha\lambda_z]}{\xi^2(\bar{s})-2\xi(\bar{s})+\alpha^2} + \alpha [\lambda_z-\alpha\varepsilon\sin^2\theta(\bar{s})] \right\}d\bar{s} \nonumber \\
    & & +\int_{0}^{s}\frac{ 2\xi(\bar{s}) \epsilon_r \sqrt{\tilde R(\bar{s})}}{\xi^2(\bar{s})-2\xi(\bar{s})+\alpha^2}d\bar{s} \nonumber \\
    & = & N_{\text{BL}}(s) + N_{\xi}(s).
\end{eqnarray}
Here
\begin{eqnarray}
    N_{\text{BL}}(s) & := &\int_{0}^{s} \left\{ \frac{[\xi^2(\bar{s})+\alpha^2][(\xi^2(\bar{s})+\alpha^2)\varepsilon-\alpha\lambda_z]}{\xi^2(\bar{s})-2\xi(\bar{s})+\alpha^2}+\alpha(\lambda_z-\alpha\varepsilon\sin^2\theta(\bar{s})) \right\} d\bar{s} \nonumber \\
    & = & \tilde N_\mathrm{BL}(s) - \alpha^2 \varepsilon N_\theta(s),
\end{eqnarray}
where
\begin{eqnarray}
    \tilde N_\mathrm{BL}(s) & := & \int_0^s \frac{\left[ \xi^2(\bar s) + \alpha^2 \right]^2 \varepsilon - 2 \alpha \lambda_z \xi(\bar s)}{\xi^2(\bar s) - 2 \xi(\bar s) + \alpha^2} d \bar s, \\
    N_\theta(s) & := & \int_0^s \sin^2 \theta(\bar s) d \bar s,
\end{eqnarray}
and 
\begin{equation}
    N_{\xi}(s) := 2\int_{0}^{s}\frac{\xi(\bar{s}) \epsilon_r \sqrt{\tilde R(\bar{s})}}{\xi^2(\bar{s})-2\xi(\bar{s})+\alpha^2}d\bar{s}.
\label{eq:Nksi}
\end{equation}
The integrals $\tilde N_\mathrm{BL}(s)$ and $N_\theta(s)$ were computed in \cite{CHM2023}.

As before, we start by evaluating the integral in Eq.\ (\ref{eq:Nksi}), which we write in the form
\begin{equation}
      N_{\xi}(s)= 2\int_{0}^{s}\frac{\xi(\bar{s})}{\xi^2(\bar{s})-2\xi(\bar{s})+\alpha^2} \frac{d \xi(\bar s)}{d \bar s} d\bar{s}.
\end{equation}
Changing the integration variable form $s$ to $\xi=\xi(s)$ we obtain
\begin{equation}
\label{Nxisol}
    N_\xi (s) = 2 \int_{\xi_0}^{\xi(s)} \frac{\bar{\xi}}{\bar{\xi}^2-2\bar{\xi}+\alpha^2}d\bar{\xi} = \log \left[ \frac{\xi^2(s) - 2 \xi(s) +\alpha^2}{\xi_0^2-2 \xi_0+\alpha^2} \right] + 2 \left[\frac{ \arctan \left(\frac{\xi(s) - 1}{\sqrt{\alpha^2 - 1}}\right)}{\sqrt{\alpha^2 -1}}- \frac{ \arctan \left(\frac{\xi_0 - 1}{\sqrt{\alpha^2 -1}}\right)}{\sqrt{\alpha^2 -1}} \right],
\end{equation}
which again together with Eq.\ (\ref{xisol}) provides the solution.

A regularization of the sum $\tilde N_\mathrm{BL}(s) + N_\xi(s)$ can be done in many ways. One of them is, however, particularly convenient. Note that
\begin{equation}
    \frac{(\xi^2 + \alpha^2)^2 \varepsilon - 2 \alpha \lambda_z \xi}{\xi^2 - 2 \xi + \alpha^2} = (\xi^2 + \alpha^2) \varepsilon + \frac{2 \xi \left[ (\xi^2 + \alpha^2) \varepsilon - \alpha \lambda_z \right]}{\xi^2 - 2 \xi + \alpha^2}.
\end{equation}
Therefore,
\begin{eqnarray}
    \tilde N_\mathrm{BL}(s) + N_\xi(s) & = & \int_0^s \left\{ \frac{2 \xi(\bar s) \left[ (\xi^2(\bar s) + \alpha^2) \varepsilon - \alpha \lambda_z + d\xi(\bar s) / d\bar s \right]}{\xi^2(\bar s) - 2 \xi(\bar s) + \alpha^2} + [\xi^2(\bar s) + \alpha^2] \varepsilon \right\} d \bar s \nonumber \\
    & = & \int_0^s \left\{ \frac{2 \xi(\bar s) \left[ \delta_1 \xi^2(\bar s) + \kappa \right]}{[\xi^2(\bar s) + \alpha^2] \varepsilon - \alpha \lambda_z - d \xi(\bar s) / d \bar s} + [\xi^2(\bar s) + \alpha^2] \varepsilon \right\} d \bar s.
    \label{regularN}
\end{eqnarray}
The advantage of this choice is that now a potentially problematic denominator in Eq.\ (\ref{regularN}) has the same form as in Eq.\ (\ref{regularJ}).

\subsection{Regularity at horizons}
\label{sec:regularity}

We see from preceding subsections that the regularity of the expressions for $\varphi^\prime$ and $T^\prime$ depends on the signs of
\begin{equation}
    A := (\xi^2 + \alpha^2) \varepsilon - \alpha \lambda_z
\end{equation}
and $\epsilon_r$. It can be shown that for timelike future-directed geodesics at $\xi > \xi_+$ (in Boyer--Lindquist Block I), one has $A > 0$. Thus, by continuity, for $\epsilon_r = -1$ (incoming geodesics) and $\tilde R > 0$, we have $A - d \xi(s)/ds = A - \epsilon_r \sqrt{\tilde R} > 0$ up to the horizon at $\xi_+$. Hence, for incoming future-directed timelike geodesics both $\varphi^\prime$
and $T^\prime$ remain regular at the event horizon joining Blocks I and II.

A proof that for a timelike future-directed geodesic at $\xi > \xi_+$, one must have $A > 0$ can be found in \cite{Rioseco2024}, but we repeat it here for completeness. Note first that the vector
\begin{equation}
    N = \left( 1 + \frac{2 M r}{\rho^2} \right) \partial_{t^\prime} - \frac{2 M r}{\rho^2} \partial_r
\end{equation}
is normal to hypersurfaces of constant time $t^\prime$. Lowering the indices in $N$ one gets $N_\mu = (-1,0,0,0)$. Since $g(N,N) = - (1 + 2 M r/\rho^2)$, it is also timelike, as long as $r > 0$. This vector defines a time orientation. Consider a vector
\begin{equation}
    X := (r^2 + a^2) \partial_{t^\prime} + a \partial_{\varphi^\prime}.
\end{equation}
O'Neill refers to $X$ as one of ``the canonical Kerr vector fields'' (\cite{Neill1995}, p.\ 60).
It satisfies $g(X,X) = - \rho^2 \Delta$, and thus it is timelike for $\Delta > 0$, i.e., for $r < r_-$ or $r > r_+$ (in Blocks I and III). It is also future-directed, since $g(N,X)= - dt^\prime(X) = -(r^2 + a^2) < 0$. On the other hand,
\begin{equation}
    p(X) = (r^2 + a^2) p_{t^\prime} + a p_{\varphi^\prime} = - \left[ (r^2 + a^2) E - a l_z \right] = - M^2 m \left[ (\xi^2 + \alpha^2) \varepsilon - \alpha \lambda_z \right],
\end{equation}
where $p$ denotes the momentum covector. A future-directed momentum $p$ has to satisfy $p(X) < 0$, and hence $A = (\xi^2 + \alpha^2) \varepsilon - \alpha \lambda_z > 0$. We emphasise that negative values of $\varepsilon$ are still allowed within the ergosphere.

In Block II, where $\Delta < 0$, i.e., for $r_- < r < r_+$, Rioseco and Sarbach \cite{Rioseco2024} propose to use the vector
\begin{equation}
    Y := 2 M r \partial_{t^\prime} + \Delta \partial_r + a \partial_{\varphi^\prime}.
\end{equation}
It satisfies $g(Y,Y) = \rho^2 \Delta$ and, consequently, it is timelike for $r_- < r < r_+$. Since $g(N,Y) = -dt^\prime(Y) = - 2 M r$, it is also future-directed in $r_- < r < r_+$. On the other hand,
\begin{equation}
    p(Y) = 2 M r p_{t^\prime} + \Delta p_r + a p_{\varphi^\prime} = - 2 M r E + \Delta p_r + a l_z.
\end{equation}
Since $\Delta p_r = \epsilon_r \sqrt{R} + 2 M r E - a l_z$, we get $p(Y) = \epsilon_r \sqrt{R}$. Thus in Block II, the momentum $p$ can be future-directed only for $\epsilon_r = -1$.

Turning to the ``regularization'' applied in Eqs.\ (\ref{regularJ}) and (\ref{regularN}) note that it is based on the observation that
\begin{equation}
    A^2 - \tilde R = (\xi^2 - 2 \xi + \alpha^2)(\delta_1 \xi^2 + \kappa),
\end{equation}
and hence $A^2 - \tilde R$ has two real zeros, precisely at $\xi = \xi_\pm$. On the other hand
\begin{equation}
    A^2 - \tilde R = \left( A - \epsilon_r \sqrt{\tilde R} \right) \left( A + \epsilon_r \sqrt{\tilde R} \right) = \left( A - \frac{d \xi}{ds} \right) \left(A + \frac{d \xi}{ds} \right).
\end{equation}
Thus, if $A - \epsilon_r \sqrt{\tilde R}$ remains non-zero at a given radius $\xi = \xi_+$ or $\xi = \xi_-$, then $A + \epsilon_r \sqrt{\tilde R}$ must be zero for the same radius, and \textit{vice versa}.  In other words, horizon-penetrating Kerr coordinates do not allow for a description in which a given trajectory can cross smoothly horizons at $\xi = \xi_+$ or $\xi = \xi_-$ in both radial directions $\epsilon_r = \pm 1$ at the same time. This is, of course, consistent with the behavior shown in Fig.\ \ref{fig:penrose}. Suppose that a given trajectory passes from Block I to Block II, and then to Block III, where it encounters a radial turning point and continues further with $\epsilon_r = +1$. Such a trajectory will hit the Cauchy horizon at $\xi = \xi_-$ with $\epsilon_r = +1$, where both $T^\prime$ and $\varphi^\prime$ would diverge. We show this behavior in particular examples in Sec.\ \ref{sec:examples}. It would also be tempting to illustrate this situation directly in Penrose diagrams in Fig.\ \ref{fig:penrose}, which are however plotted for points at the axis, and assuming that $d \theta = d \varphi = 0$. An illustration of this kind should be possible in terms of projection diagrams defined in \cite{Chrusciel2012}. Unfortunately, from the computational point of view, such diagrams are much more difficult to draw exactly.

\section{Examples}
\label{sec:examples}

\begin{table}[t]
\caption{\label{tab:wyniki} Parameters of geodesics shown in Figs.\ \ref{fig:tr1} to \ref{fig:additional}. In all examples $\alpha = 0.8$, $\theta_0 = 0.85$, $\varphi_0^\prime = 0.33$.}
\begin{ruledtabular}
\begin{tabular}{ccccccccc}
     Fig.\ No.\ & $\delta _1$ &  $\varepsilon ^2$  & $\kappa$  &  $\lambda_z$  &  $\alpha$  &  Real zeros of $\tilde R(\xi)$&  Real zeros of $\tilde \Theta(\theta)$& $\xi _0$ \\ \hline
     2 & 1 & 1.1 & 12 & $-1$ & 0.8 & $-24.3351$,\, 0.254136 & 2.81429,\, 0.327303 & 8 \\
     3 & 1 & 0.95  & 12 & 3 & 0.8 & $\begin{array}{c} 0.22019,\, 1.63896, \\ 8.44487,\, 29.696 \end{array} $& 2.29552,\, 0.846071 &   10 \\
     4 & 1 & 1.1  & 12 & 3 & 0.8 & $\begin{array}{c} -26.4861,\, 0.230431,\\ 1.67987,\, 4.57578 \end{array} $ & 2.30596,\, 0.835636 &  10 \\
     5 & 1 & 30  & 12 & $-0.05$ & 0.8 & --- &  $\begin{array}{c} 3.12646,\, 2.2675,\\ 0.874094,\, 0.0151329 \end{array} $&  10 \\
     6 & 0 & 1  & 60 & 4.47214 & 0.8 & $\begin{array}{c} -8.927,\, 0.296172, \\ 1.60191,\, 7.02892 \end{array} $& 2.56341,\, 0.578185 &  10 \\
     7 & 0 & 1  & 0.6 & $-0.111803$ & 0.8 & $-0.721257$,\, $-0.137167$ & $\begin{array}{c} 2.96415,\, 2.22796,\\ 0.913634,\, 0.177447 \end{array}$ & 10 \\
     8 & 0 & 1  & 0.4 & $-0.00912871$ & 0.8 & --- &$\begin{array}{c}  3.12688,\, 2.25351,\\ 0.88808,\, 0.0147079 \end{array}$ & 10 \\
     9 & 1 & 1.1   & 10 & $-0.2$ & 0.8 & $-3.59963$,\, 0.329668 & 3.0772,\, 0.0643885 & 12
\end{tabular}
\end{ruledtabular}
\end{table}

In this section we discuss a collection of sample solutions obtained with the help of formulas derived in preceding sections.

We perform our computations using \textit{Wolfram Mathematica} \cite{wolfram}. The formulas for $\xi(s)$ and $\mu(s)$, or equivalently $\theta(s)$, can be encoded directly. The formulas for the azimuthal angle $\varphi^\prime(s)$ and the time $T^\prime(s)$ require a regularization, if a geodesic crosses one of the horizons. In practice, we substitute in Eq.\ (\ref{regularJ}) expressions for $\xi(s)$ and $\theta(s)$ given by Eqs.\ (\ref{xisol}) and (\ref{musol}), and evaluate the resulting integral numerically. In principle, the derivative $d \xi(s)/ds$ in Eq.\ (\ref{regularJ}) could be expressed as $d \xi(s)/ds = \epsilon_r \sqrt{\tilde R}$, by virtue of Eq.\ (\ref{eq:mot.ksi}). Using this form turns out to be problematic, as it requires knowledge of the sign $\epsilon_r$, which changes at the radial turning points. To circumvent this difficulty, we use in Eq.\ (\ref{regularJ}) the derivative $d \xi(s)/ds$ obtained by a direct differentiation of Eq.\ (\ref{xisol}). Although the integral (\ref{regularJ}) can be evaluated analytically, similarly to the calculation presented in \cite{CHM2023}, we find computing it numerically to be quite effective. In a sense, we try to combine in our implementation the best of the two approaches---elegant formulas for the radial and polar coordinates and relatively straightforward numerical integrals providing $\varphi^\prime$ and the time coordinate $T^\prime$.

\begin{figure}
\begin{center}
\includegraphics[width=0.4\textwidth]{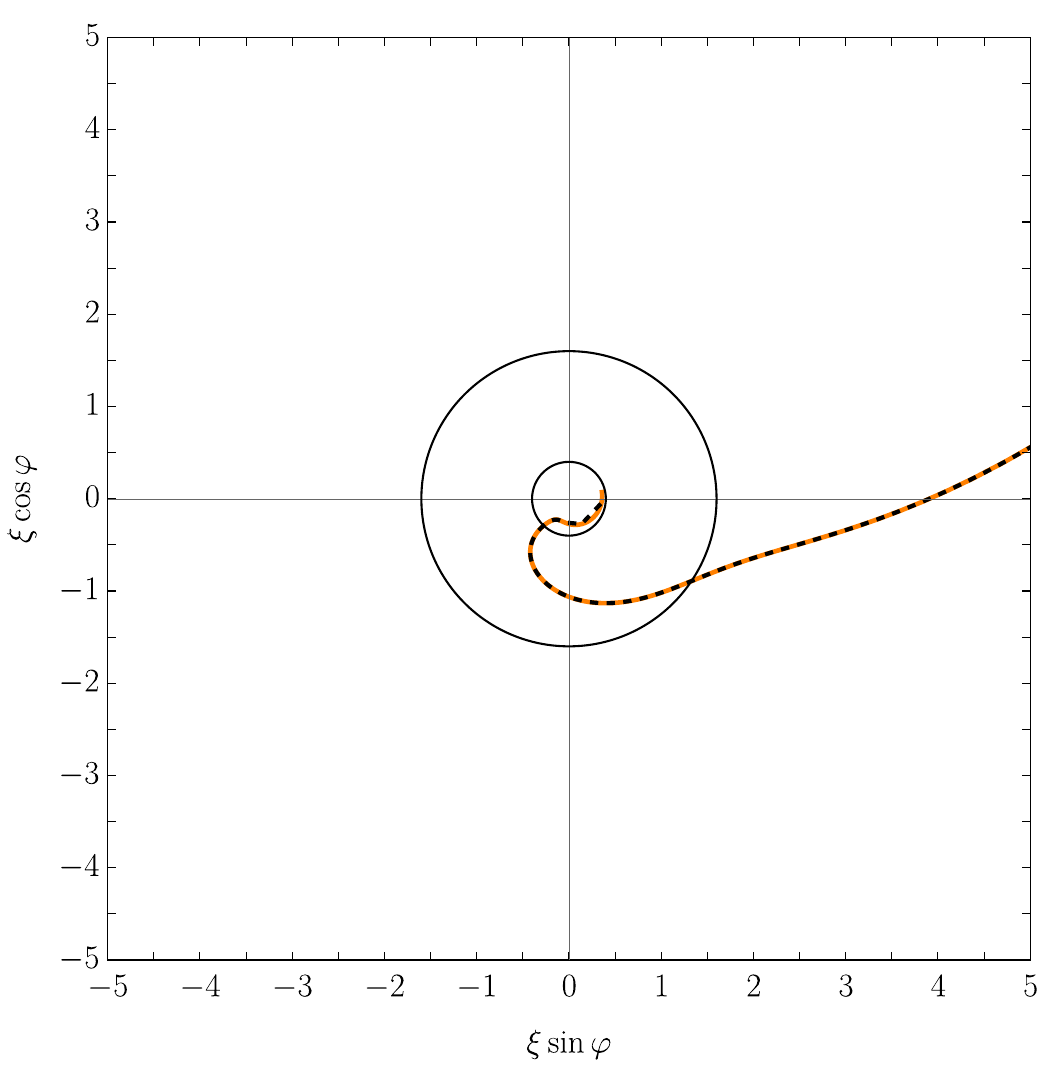}
\includegraphics[width=0.4\textwidth]{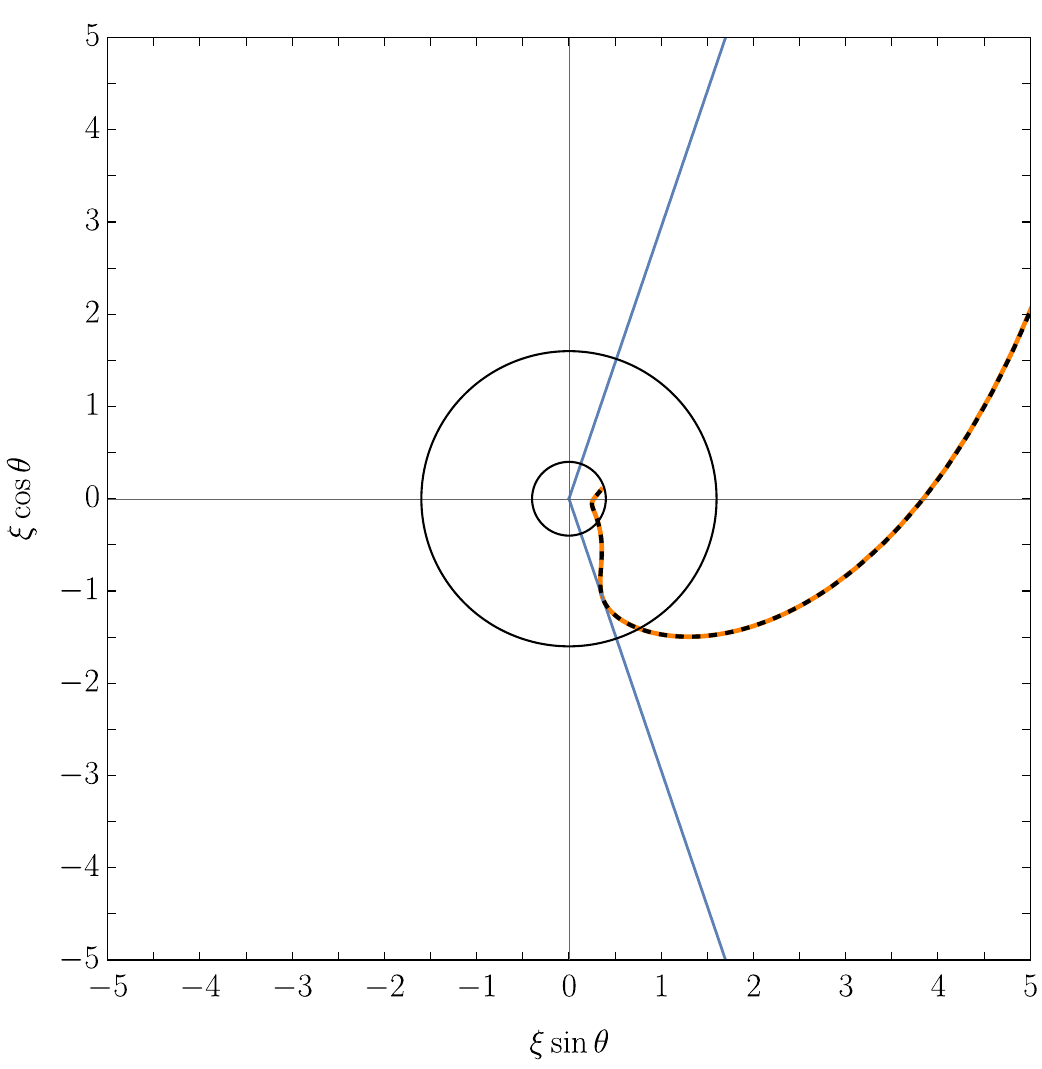}\\
\includegraphics [width=0.4\textwidth]{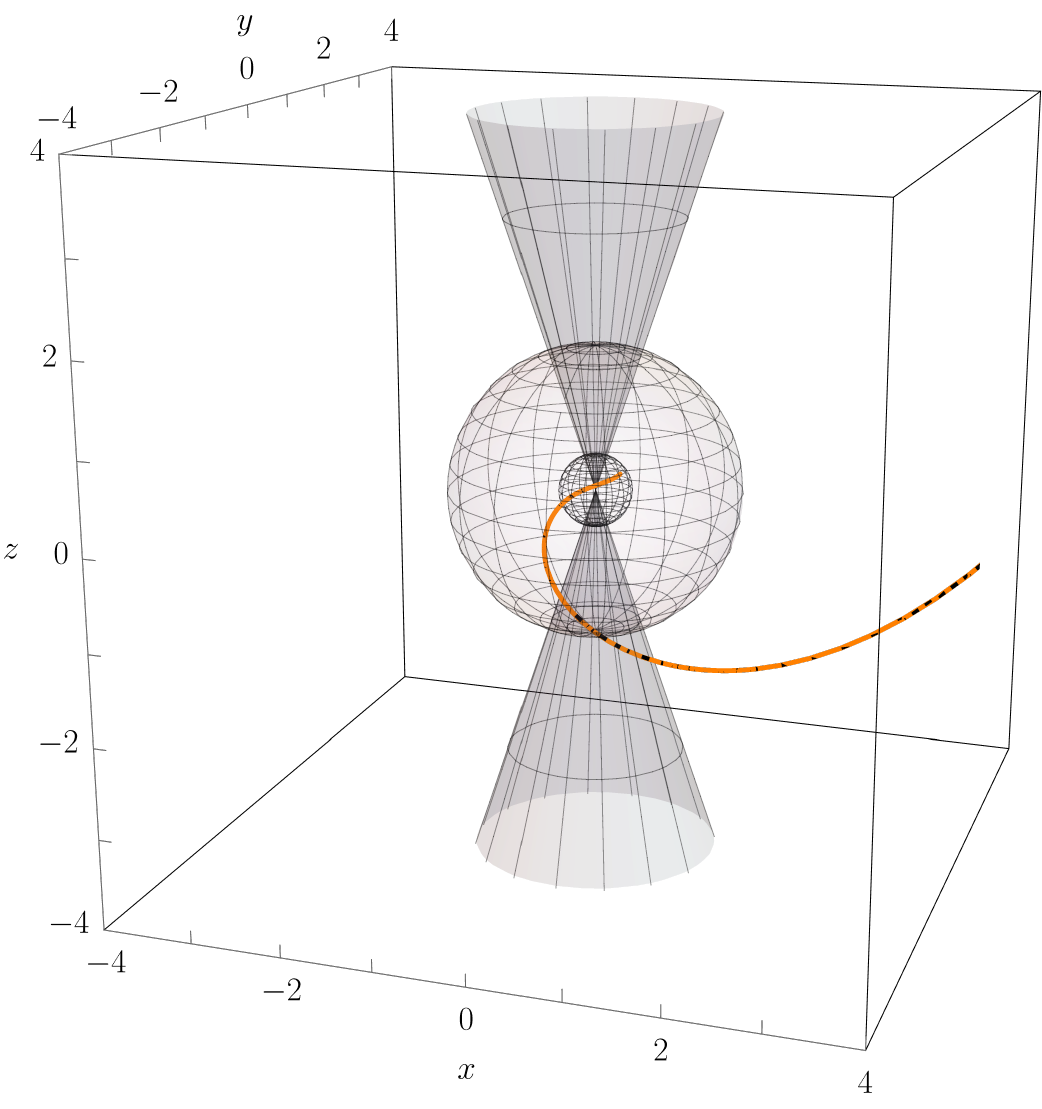}
\end{center}
\caption{\label{fig:tr1} A timelike orbit with $\varepsilon^2 = 1.1$, $\lambda_z = -1$, $\alpha = 0.8$, $\kappa = 12$ and initial position at $\xi_0 = 8$, $\theta_0 = 0.85$, $\epsilon_{r,0} = -1$, $\epsilon_{\theta,0} = 1$.}
\end{figure}

\begin{figure}
\begin{center}
\includegraphics[width=0.4\textwidth]{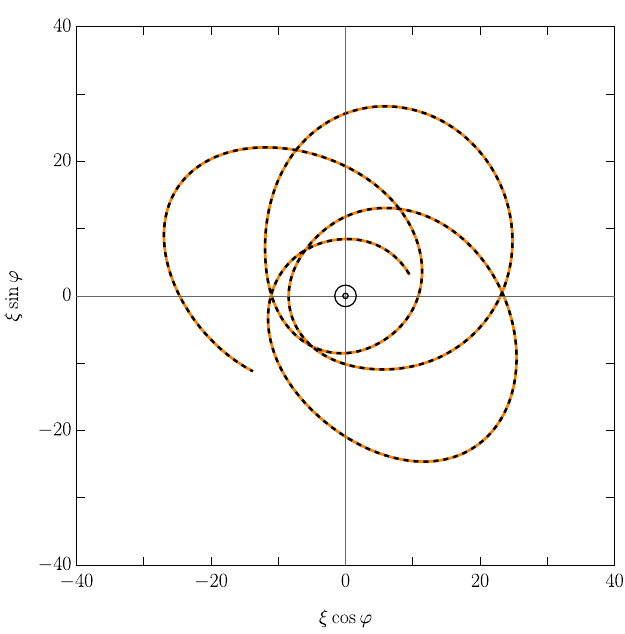}
\includegraphics[width=0.4\textwidth]{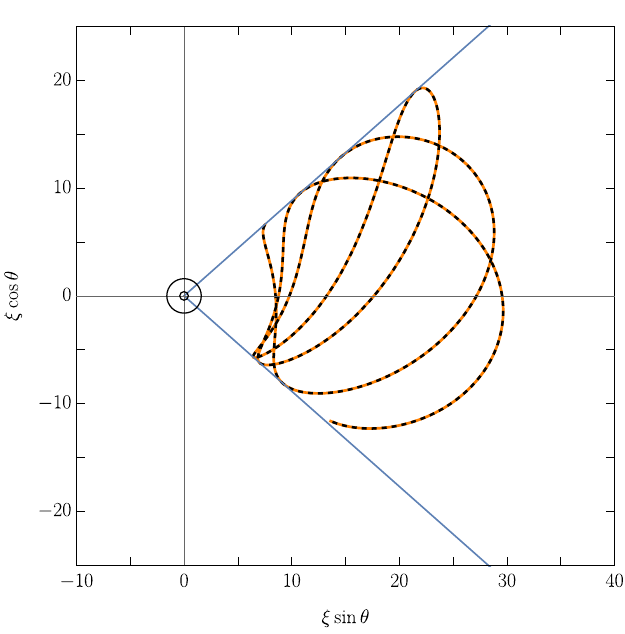}\\
\includegraphics[width=0.4\textwidth]{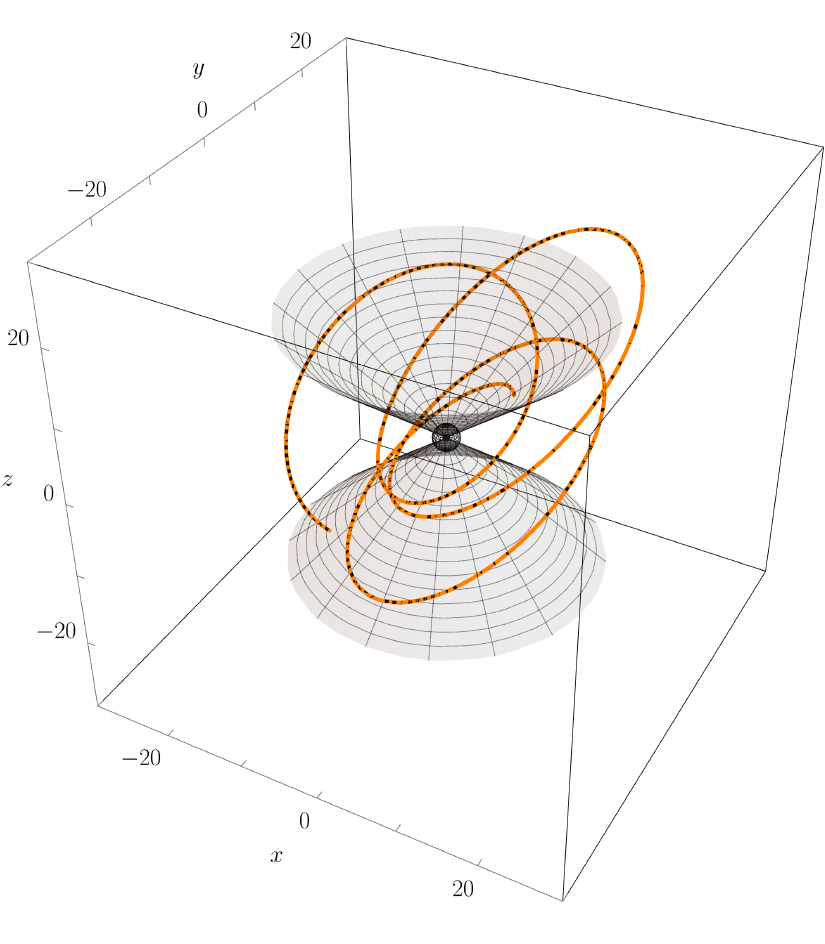}
\end{center}
\caption{\label{fig:tr2} Same as in Fig.\ \ref{fig:tr1}, but for a timelike bound orbit with $\varepsilon^2 = 0.95$, $\lambda_z = 3$, $\alpha = 0.8$, $\kappa = 12$, and an initial position at $\xi_0 = 10$, $\theta_0 = 0.85$, $\epsilon_{r,0} = -1$, $\epsilon_{\theta,0} = 1$.}
\end{figure}

\begin{figure}
\begin{center}
\includegraphics[width=0.4\textwidth]{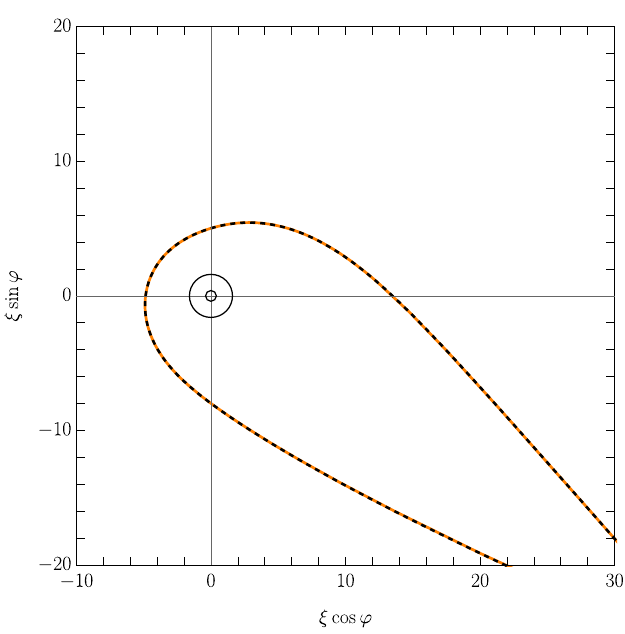}
\includegraphics[width=0.4\textwidth]{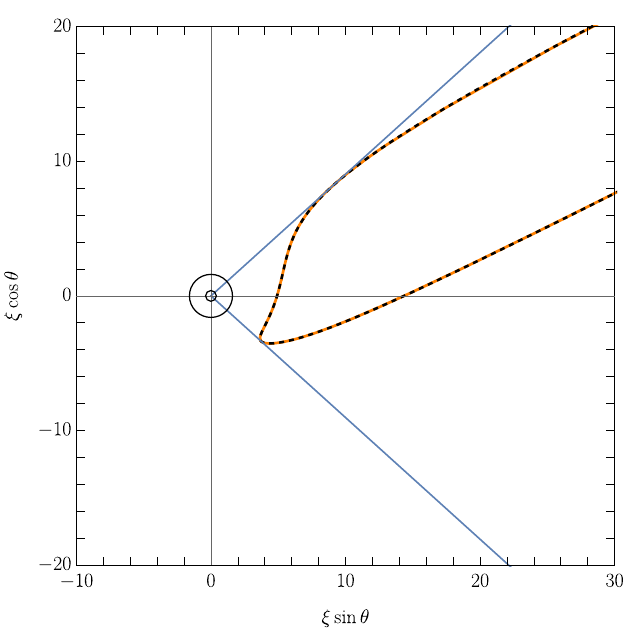}\\
\includegraphics[width=0.4\textwidth]{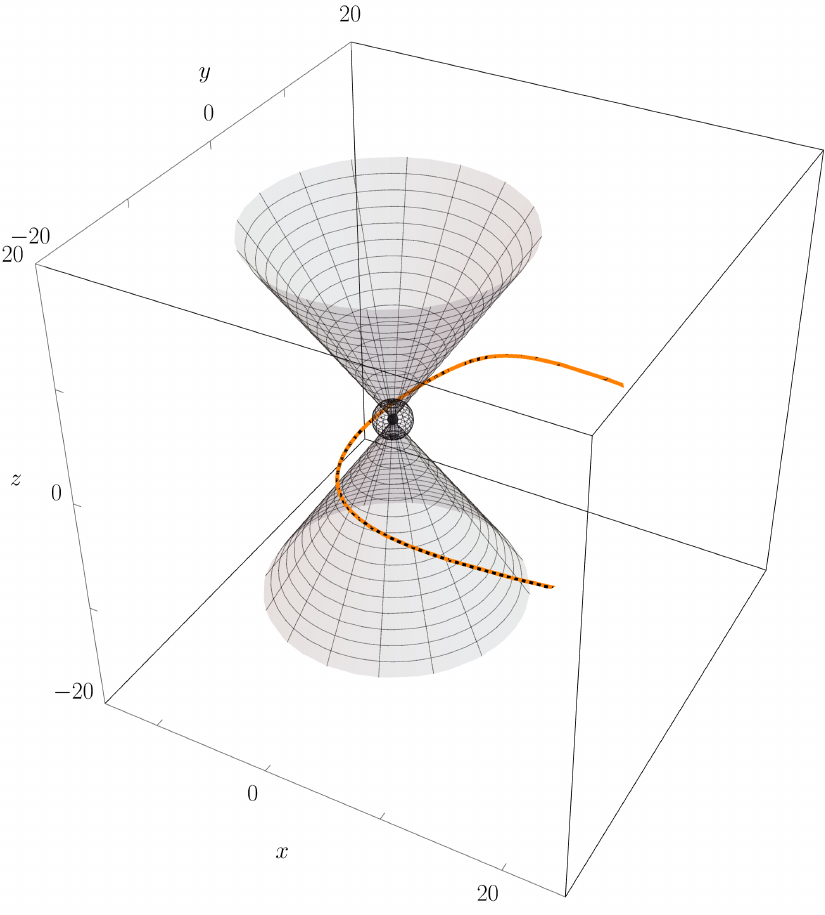}
\end{center}
\caption{\label{fig:tr4} Same as in Fig.\ \ref{fig:tr1}, but for a timelike unbound scattered orbit with $\varepsilon^2 = 1.1$, $\lambda_z = 3$, $\alpha = 0.8$, $\kappa = 12$, and an initial position at $\xi_0 = 10$, $\theta_0 = 0.85$, $\epsilon_{r,0} = -1$, $\epsilon_{\theta,0} = 1$.}
\end{figure}

\begin{figure}
\begin{center}
\includegraphics[width=0.4\textwidth]{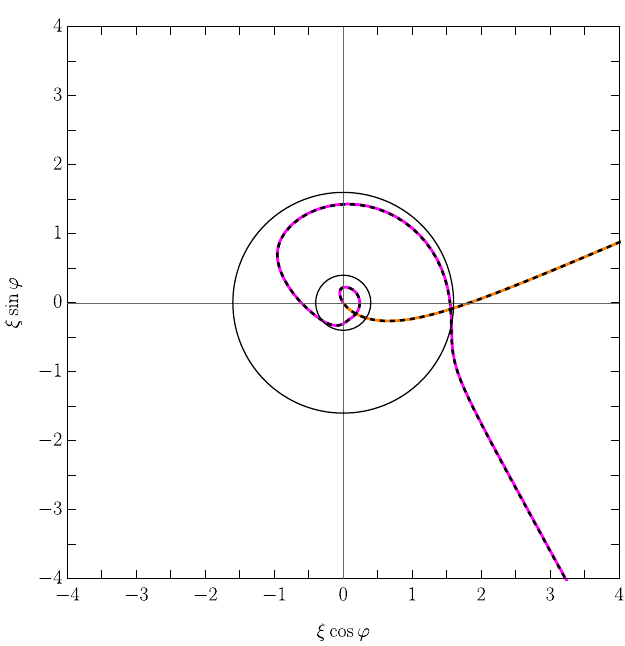}
\includegraphics[width=0.4\textwidth]{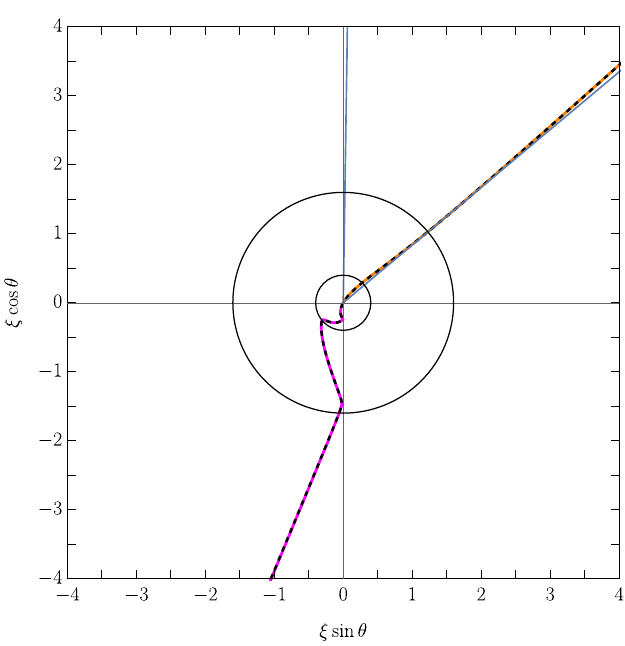}\\
\includegraphics[width=0.4\textwidth]{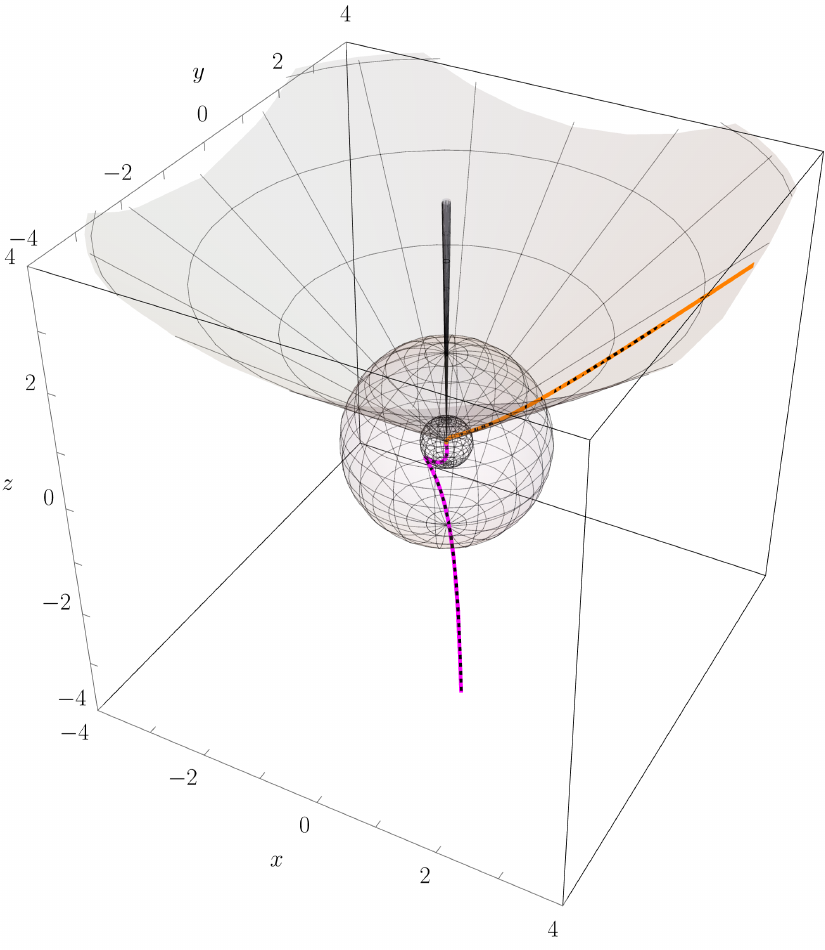}
\includegraphics[width=0.4\textwidth]{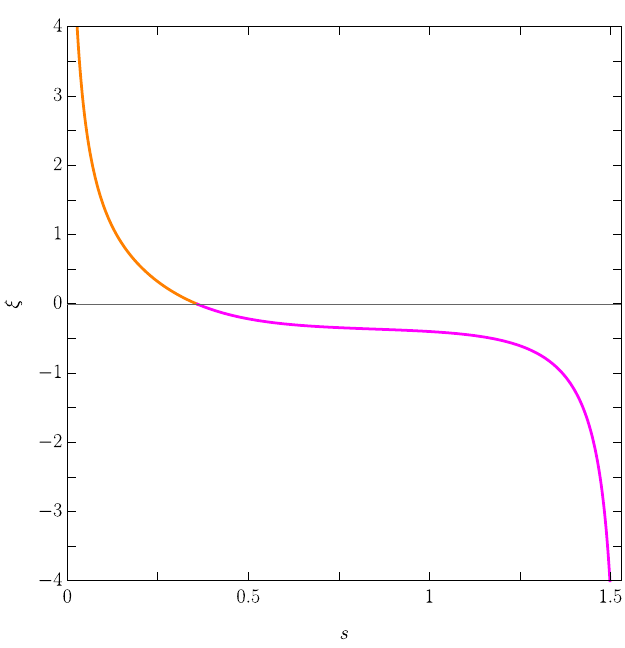}
\end{center}
\caption{\label{fig:tr5} A timelike orbit with $\varepsilon^2 = 30$, $\lambda_z = -0.05$, $\alpha = 0.8$, $\kappa = 12$, and initial position at $\xi_0 = 10$, $\theta_0 = 0.85$, $\epsilon_{r,0} = -1$, $\epsilon_{\theta,0} = 1$. A part of the trajectory with $\xi < 0$ is marked in purple.}
\end{figure}

\begin{figure}
\begin{center}
\includegraphics[width=0.4\textwidth]{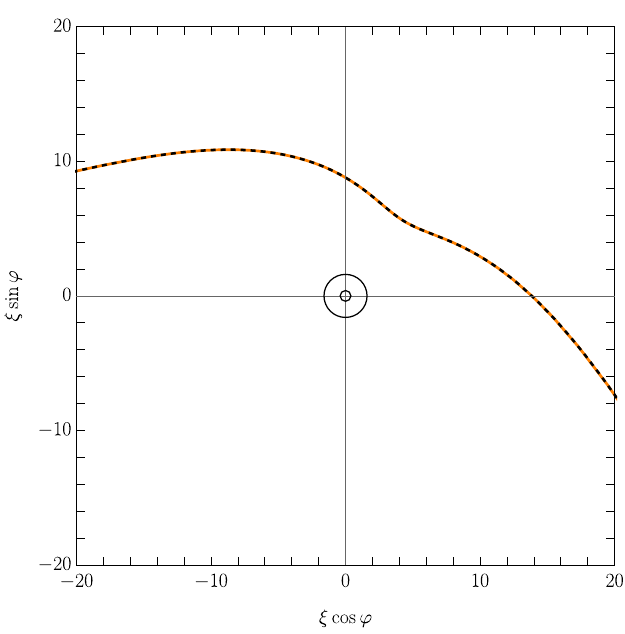}
\includegraphics[width=0.4\textwidth]{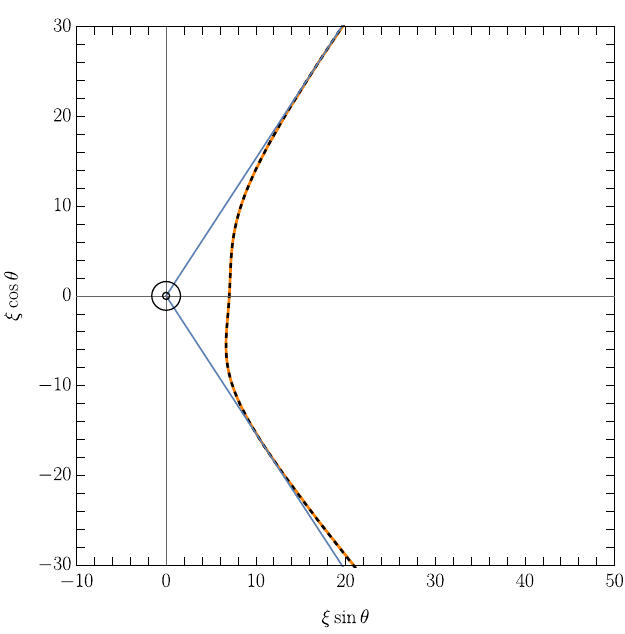}\\
\includegraphics[width=0.4\textwidth]{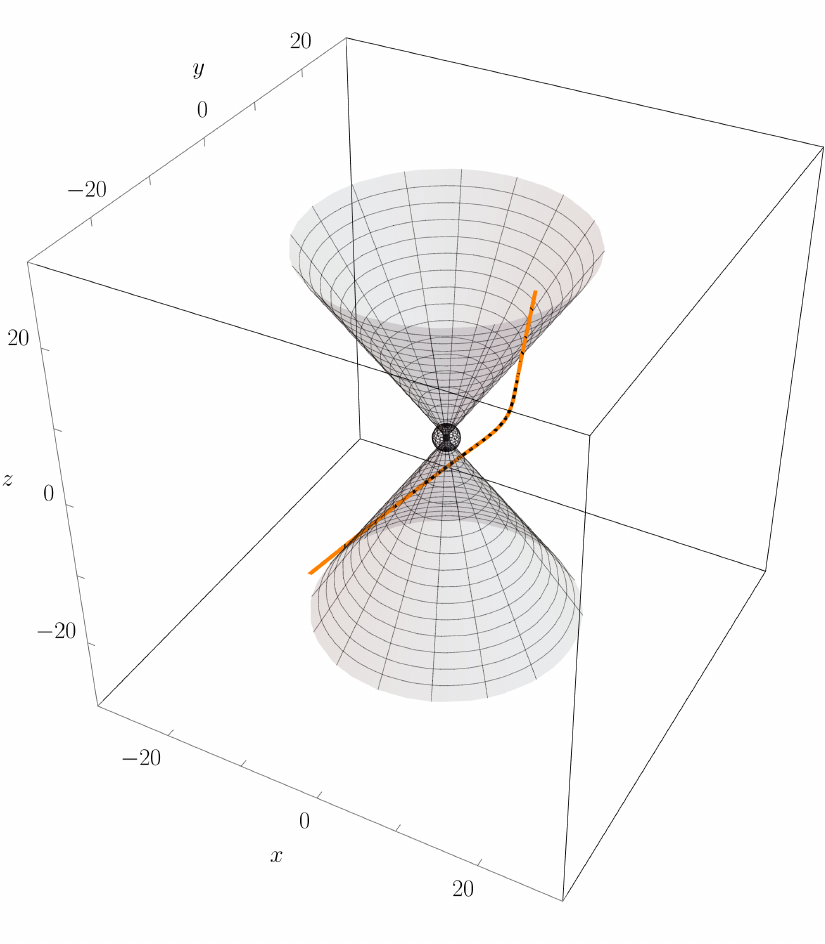}
\end{center}
\caption{\label{fig:tr7} Same as in Fig.\ \ref{fig:tr1}, but for a null unbound orbit with $\varepsilon^2 = 1$, $\lambda_z \approx 4.47214$, $\alpha = 0.8$, $\kappa = 60$, and an initial position at $\xi_0 = 10$, $\theta_0 = 0.85$, $\epsilon_{r,0} = -1$, $\epsilon_{\theta,0} = 1$.}
\end{figure}

\begin{figure}
\begin{center}
\includegraphics[width=0.4\textwidth]{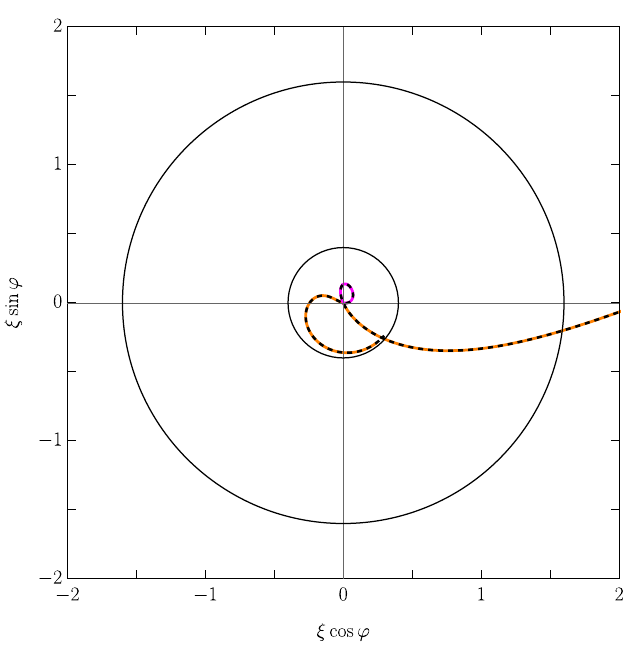}
\includegraphics[width=0.4\textwidth]{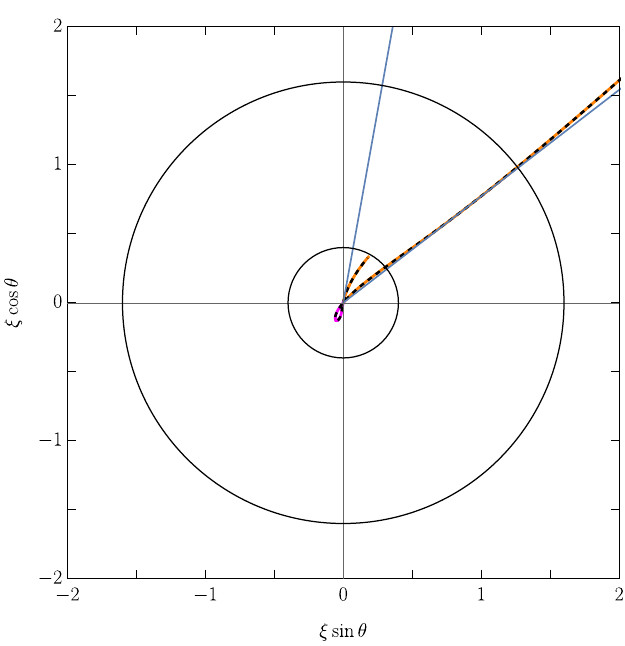}\\
\includegraphics[width=0.4\textwidth]{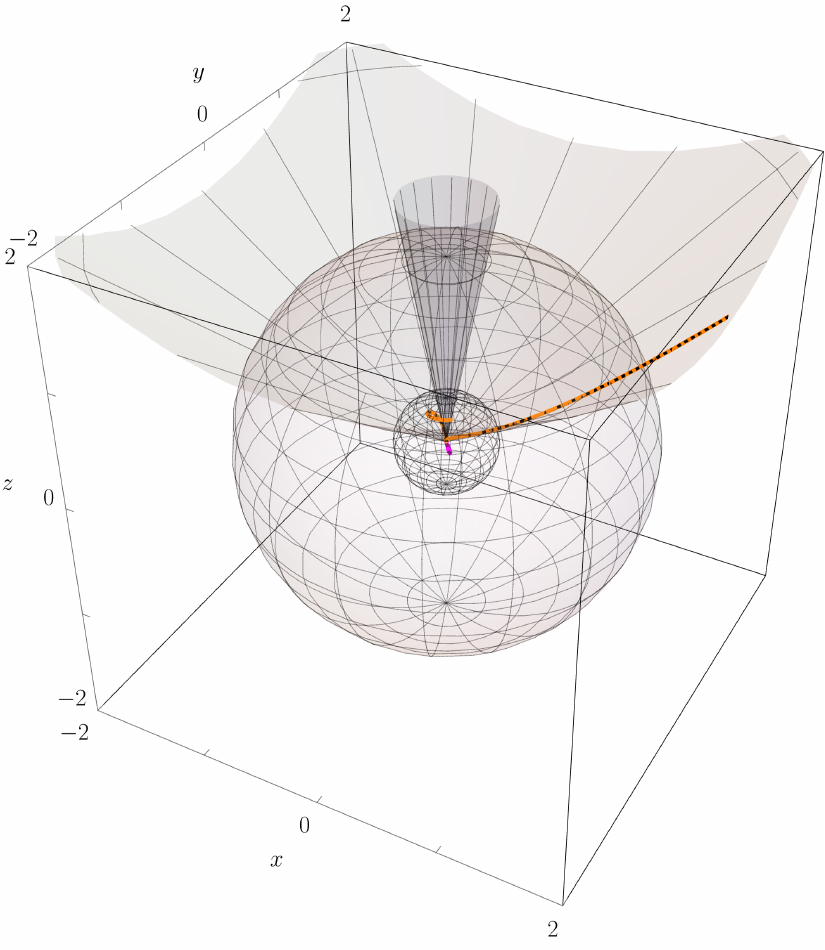}
\includegraphics[width=0.4\textwidth]{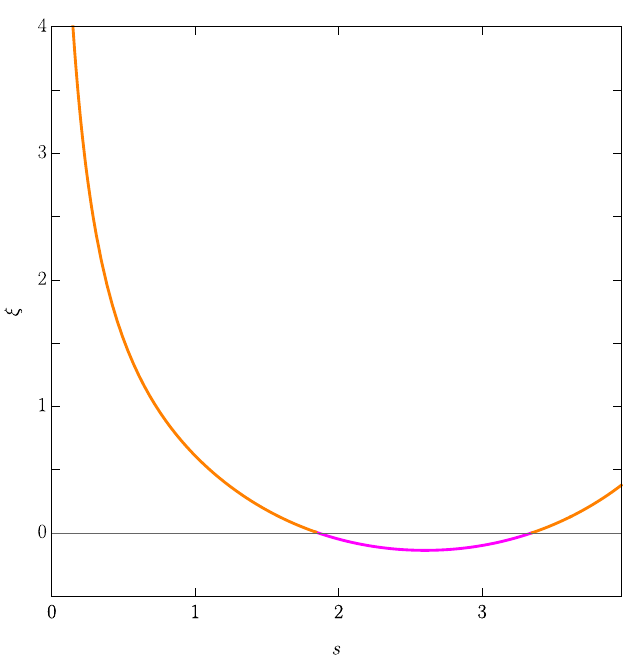}
\end{center}
\caption{\label{fig:tr9} A null orbit with $\varepsilon^2 = 1$, $\lambda_z = -0.111803$, $\alpha = 0.8$, $\kappa = 0.6$ and initial position at $\xi_0 = 10$, $\theta_0 = 0.85$, $\epsilon_{r,0} = -1$, $\epsilon_{\theta,0} = 1$.}
\end{figure}

\begin{figure}
\begin{center}
\includegraphics[width=0.4\textwidth]{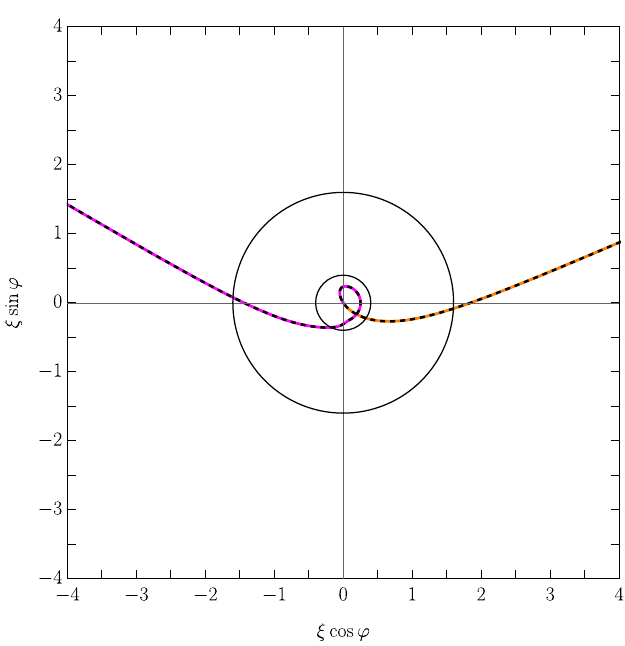}
\includegraphics[width=0.4\textwidth]{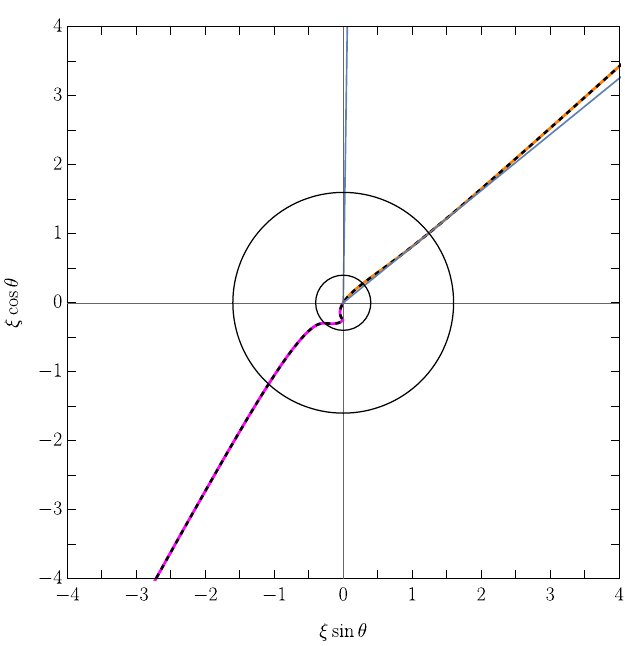}\\
\includegraphics[width=0.4\textwidth]{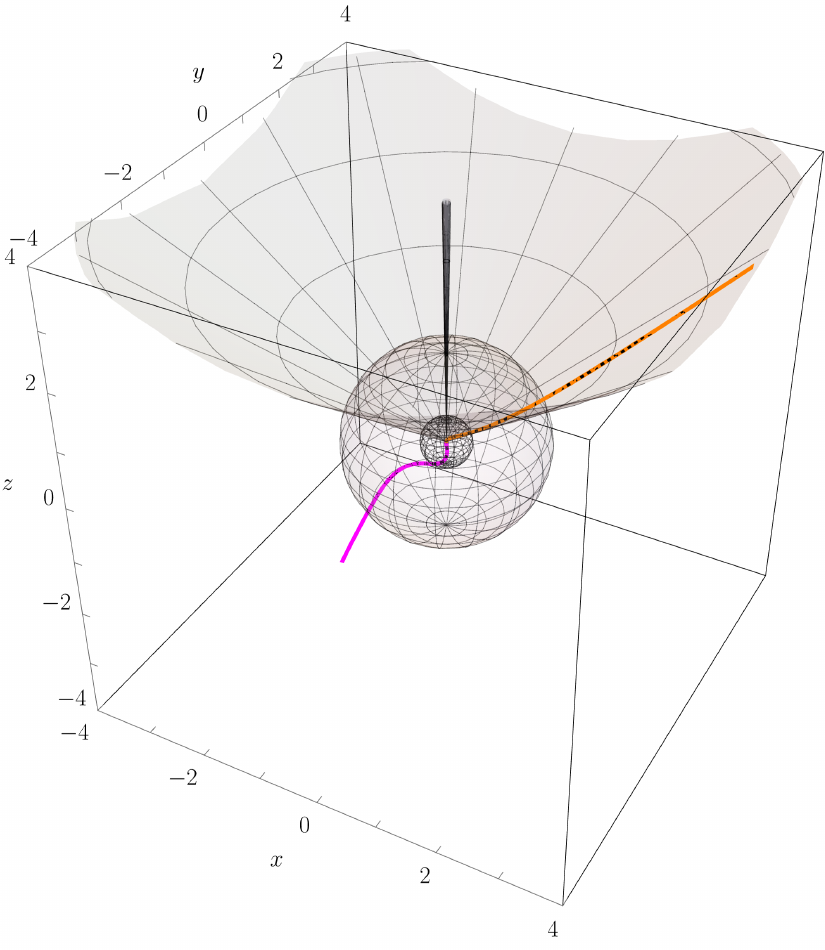}
\includegraphics[width=0.4\textwidth]{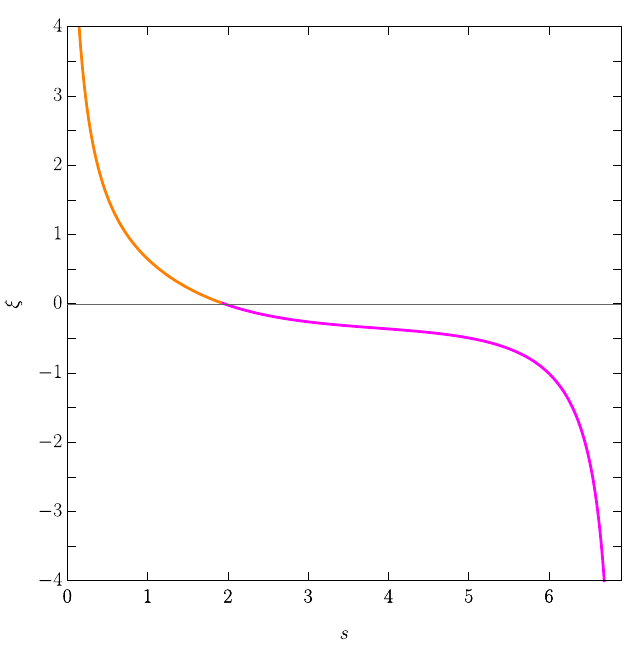}
\end{center}
\caption{\label{fig:tr10} A null orbit with $\varepsilon^2 = 1$, $\lambda_z = -0.00912871$, $\alpha=0.8$, $\kappa=0.4$, and initial position at $\xi_0 = 10$, $\theta_0 = 0.85$, $\epsilon_{r,0} = -1$, $\epsilon_{\theta,0} = 1$.}
\end{figure}

\begin{figure}
\begin{center}
\includegraphics[width=0.4\textwidth]{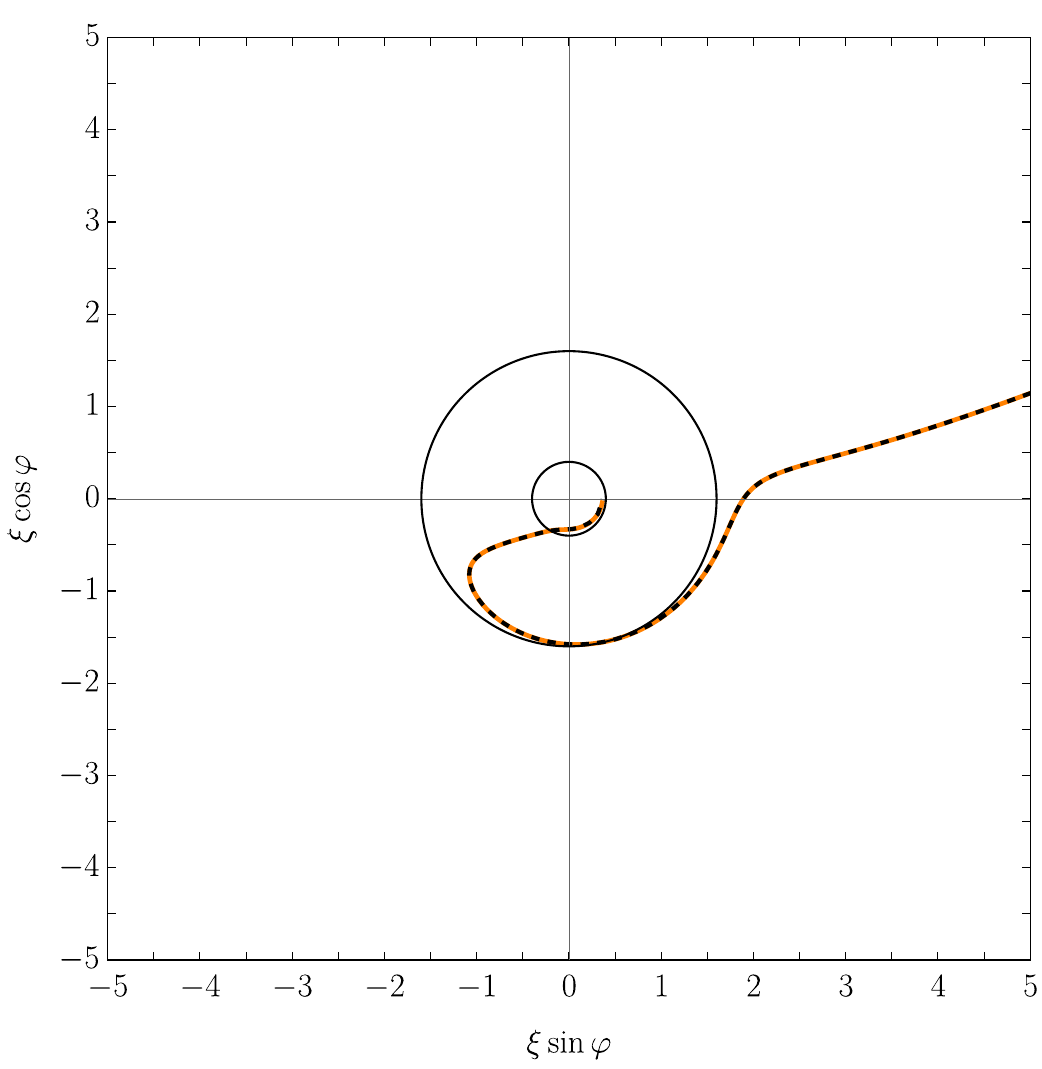}
\includegraphics[width=0.4\textwidth]{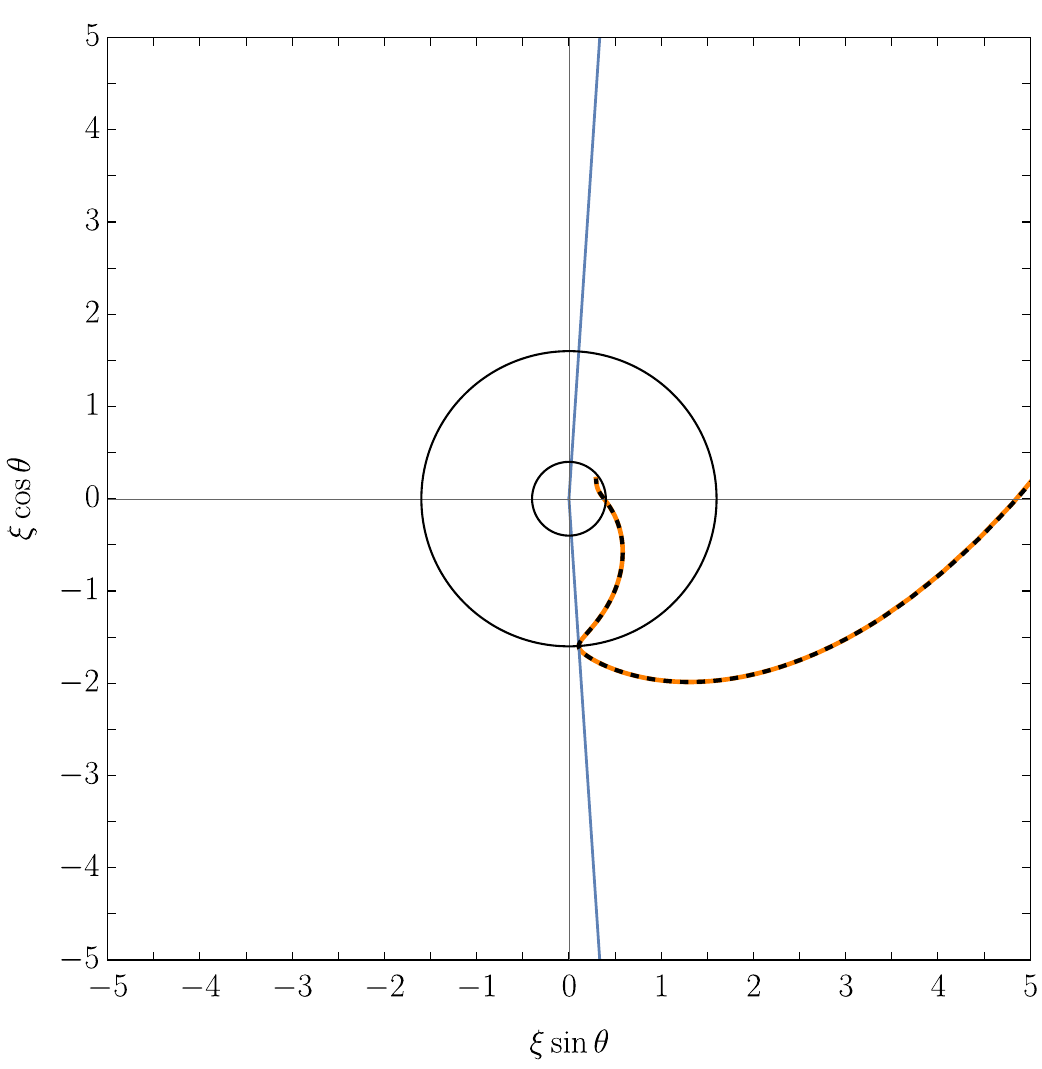}\\
\includegraphics[width=0.4\textwidth]{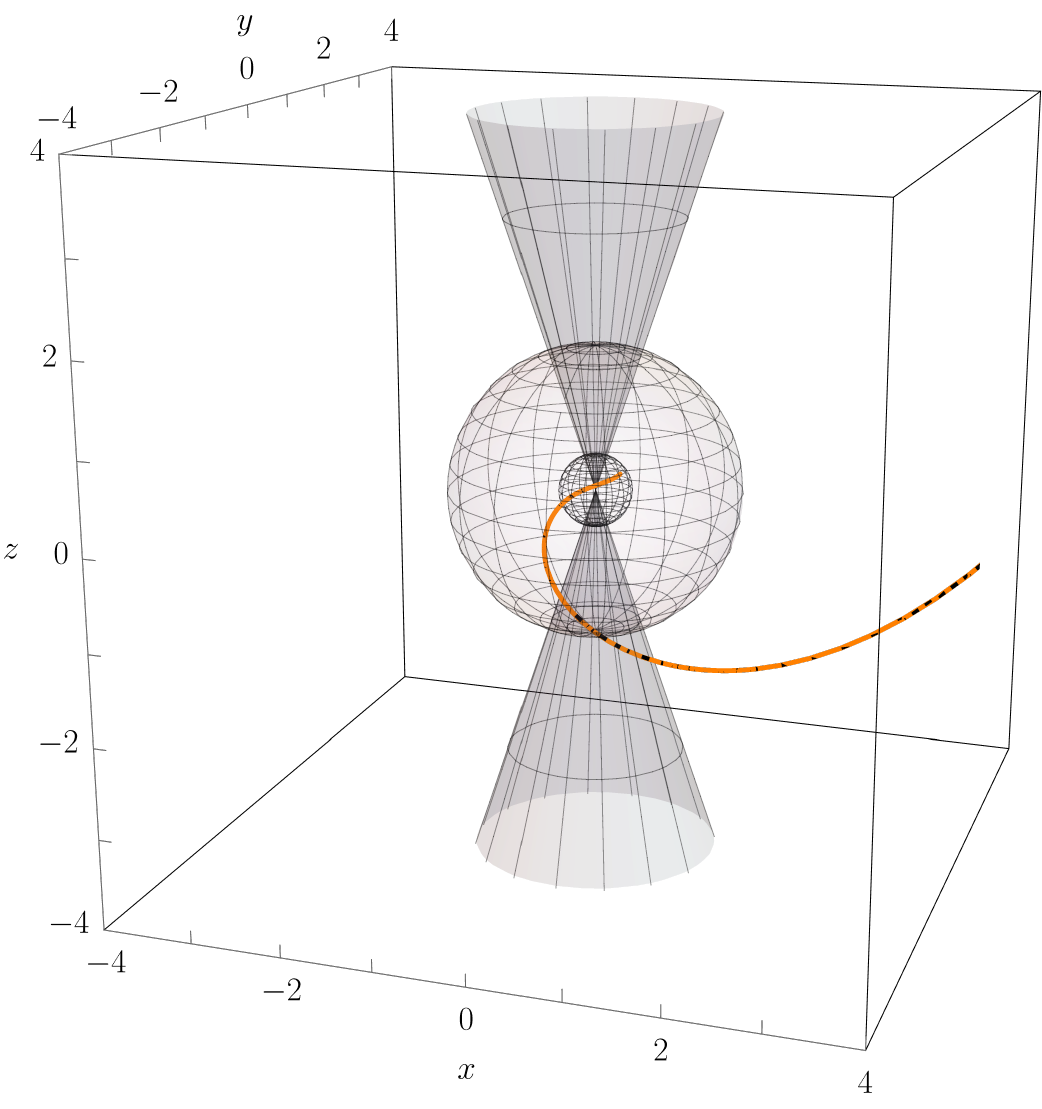}
\end{center}
\caption{\label{fig:additional} A timelike orbit with $\varepsilon^2 = 1.1$, $\lambda_z = -0.2$, $\alpha = 0.8$, $\kappa = 10$ and initial position at $\xi_0 = 12$, $\theta_0 = 0.85$, $\epsilon_{r,0} = -1$, $\epsilon_{\theta,0} = 1$.}
\end{figure}

Figures \ref{fig:tr1} to \ref{fig:additional} depict our solutions obtained for various parameters, collected in Table \ref{tab:wyniki}. They show the orbits in $\xi$--$\theta$ and $\xi$--$\varphi^\prime$ planes and in the three-dimensional space. We use Cartesian coordinates $(x,y,z)$ defined as 
\begin{subequations}
\label{cartesian}
\begin{eqnarray}
        x & = &  \xi \cos \varphi^\prime \sin \theta, \\
        y & = & \xi \sin \varphi^\prime \sin \theta, \\
        z & = & \xi \cos \theta.
\end{eqnarray}
\end{subequations}
In Table \ref{tab:wyniki} we provide also real zeros of $\tilde R(\xi)$, corresponding to radial turning points, as well as zeros of $\tilde \Theta(\theta)$. The latter define the angular ranges available for the motion along a given geodesic. We illustrate these ranges by drawing appropriate cones in three-dimensional plots or lines in $\xi$--$\theta$ plane plots. Kerr horizons at $\xi = \xi_\pm$ are depicted as spheres or circles. Finally, as a double-check of our results, we plot solutions obtained by solving numerically the Kerr geodesic equations. These numerical solutions are depicted with dotted lines. For simplicity, we omit the prime in $\varphi^\prime$ in labels of all figures in this paper.

Figure \ref{fig:tr1} shows an unbound timelike trajectory, plunging into the black hole. The trajectory crosses smoothly both horizons at $\xi = \xi_\pm$ and encounters a radial turning point located at $0 < \xi < \xi_-$. The solution for $\varphi^\prime$ can be continued smoothly up to the Cauchy horizon at $\xi = \xi_-$, where it diverges.

Figure \ref{fig:tr2} shows a standard timelike bound geodesic, corresponding to a ``Keplerian'' motion. In Fig.\ \ref{fig:tr4}, we depict an unbound timelike orbit, which does not plunge into the black hole.

A very interesting case is shown in Fig.\ \ref{fig:tr5}, depicting a timelike unbound trajectory crossing smoothly both horizons at $\xi = \xi_\pm$. This trajectory does not encounter a radial turning point and consequently it continues, in the extended Kerr spacetime, to negative radii $\xi$ (Boyer--Lindquist Block III, see also \cite{Hawking1973}, p.\ 163). In the terminology of \cite{Neill1995} such orbits are referred to as ``transits''. We illustrate the transition from $\xi > 0$ to $\xi < 0$ by plotting a segment of the geodesic corresponding to $\xi > 0$ in orange, and a segment corresponding to $\xi < 0$ in purple. Note that Cartesian coordinates defined by Eq.\ (\ref{cartesian}) allow for a change from positive to negative values of the radius, which corresponds simply to the reflection $x \to -x$, $y \to - y$, $z \to -z$, or equivalently, to a change of angular coordinates. To avoid confusion, apart from using two colors in the graphs, we also provide a plot of the radius $\xi$ versus the Mino time $s$. Since no horizons occur for $\xi < 0$, the trajectory can be continued smoothly to $\xi \to - \infty$. Note that the apparent reflection of Cartesian coordinates associated with the transition from $\xi > 0$ to $\xi < 0$ creates (in the plots) a false impression of a reflection in the polar angle $\theta$, as well as an impression that the trajectory leaves the bounds of the allowed polar motion, given by zeros of $\tilde \Theta(\theta)$. In reality, $\theta$ remains safely within the allowed range along the entire geodesic trajectory.

Figure \ref{fig:tr7} shows an example of an unbound null geodesic, scattered by the black hole. In Fig. \ref{fig:tr9}, we also deal with a null geodesic, but it is much more interesting. This trajectory crosses smoothly both horizons and continues to negative values of $\xi$, where it encounters a radial turning point. It then re-enters the region with $\xi > 0$ and can be continued up to the Cauchy horizon at $\xi = \xi_-$. In Fig.\ \ref{fig:tr9} a segment corresponding to negative radii is again marked in purple, while a segment with positive radii is plotted in orange.

Figure \ref{fig:tr10} illustrates a behavior similar to the one depicted in Fig.\ \ref{fig:tr5}, but obtained for a null geodesic. The trajectory plunges into the black hole and continues to negative radii. Figure \ref{fig:additional} shows another timelike geodesic with a behavior similar to the one illustrated in Fig.\ \ref{fig:tr1}.

\begin{figure}
\begin{center}
\includegraphics[width=0.4\textwidth]{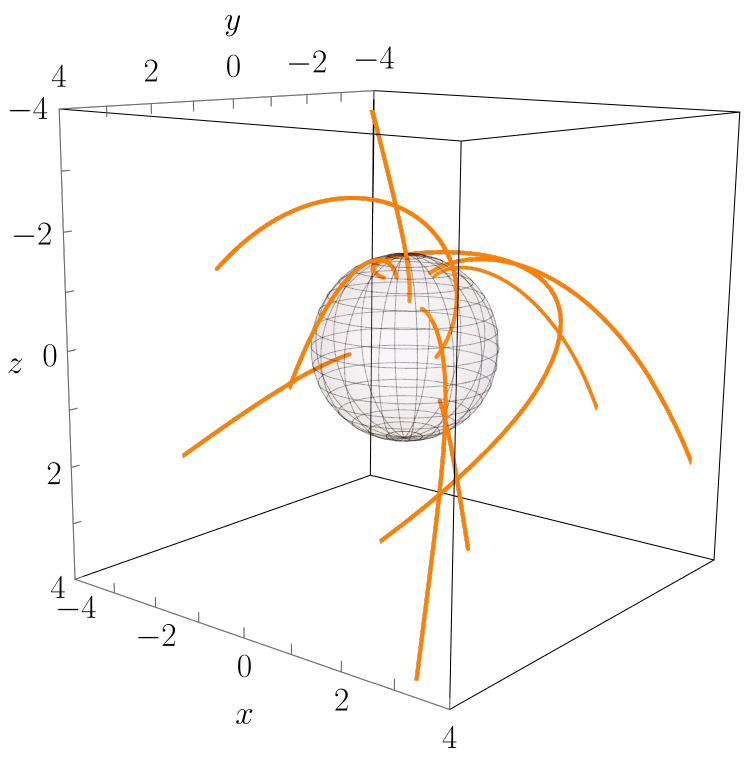}
\end{center}
\caption{\label{fig:many} A sample of timelike geodesics plunging into the black hole. The event horizon is depicted as a gray sphere. Geodesics are only plotted up to the horizon. This plot can be compared with similar ones made in Boyer--Lindquist coordinates (see, e.g., Fig.\ 6 in \cite{Dyson2023})}.
\end{figure}

Finally, in Fig.\ \ref{fig:many} we plot a collection of more or less random timelike geodesics plunging into the black hole. In this case we only show segments of geodesics located outside the black hole horizon (although they can be continued into the black hole). With this plot we aim to show that, when visualized in horizon-penetrating Kerr coordinates, the geodesics plunging into the black hole do not make an impression of swirling around the horizon---a behavior present in the Boyer--Lindquist coordinates. Our Fig.\ \ref{fig:many} should be contrasted, e.g., with Fig.\ 6 in \cite{Dyson2023}.

\section{Summary}

We provided a description of timelike and null Kerr geodesics in horizon-penetrating Kerr coordinates, extending a recent analysis of \cite{CHM2023}. From the technical point of view, the change from Boyer--Lindquist coordinates used in \cite{CHM2023} to the horizon-penetrating Kerr coordinate system only slightly affects our formalism. The radial and polar equations remain unchanged, which allows us to use relatively compact formulas (\ref{xisol}) and (\ref{musol}) for the radial and polar coordinates. Additional terms appear in equations for the azimuthal and time coordinates, but they can be integrated, once the solution to the radial equation is known. Our formulas (\ref{Jxisol}) and (\ref{Nxisol}) provide, together with Eq.\ (\ref{xisol}), all necessary expressions.

The horizon-penetrating Kerr coordinate system allows for a continuation of geodesics across horizons within the region of the extended Kerr spacetime in which it is well-defined, i.e., within Boyer--Lindquist Blocks I, II, and III (Fig.\ \ref{fig:penrose}). At the level of geodesic equations, regularity at the horizons depends explicitly on the radial direction of motion. A reasoning given in Sec.\ \ref{sec:regularity} and our examples of Sec.\ \ref{sec:examples} provide the following generic picture. A future-directed timelike or null geodesic originating outside the black hole (in Block I) and moving inward ($\epsilon_r = -1$) can encounter a radial turning point at $r > r_+$ or it can pass smoothly through the event ($r = r_+$) and the Cauchy ($r = r_-$) horizons, transiting through Block II to Block III. In Block III a trajectory may continue smoothly to $r \to - \infty$ (the so-called transit orbit), or it may get reflected at a radial turning point. In the latter case, the Kerr coordinate system only allows for a continuation up to the Cauchy horizon at $r = r_-$, where both $t^\prime$ and $\varphi^\prime$ diverge. We leave aside an obvious case of a geodesic hitting the ring singularity at $\rho^2 = 0$. It is known (see a proof in \cite{Neill1995}, p.\ 288) that a timelike or null trajectory can only hit the ring singularity, if it is located entirely within the equatorial plane.

\begin{acknowledgments}
We would like to thank Sebastian Szybka for fruitful discussions. A. C.\ and P.\ M.\ acknowledge a support of the Polish National Science Centre Grant No.\ 2017/26/A/ST2/00530.
\end{acknowledgments}

\appendix

\section{Penrose diagrams and hypersurfaces of constant Kerr time}
\label{appendix:penrose}

\begin{figure}
\begin{center}
\includegraphics[width=0.49\textwidth]{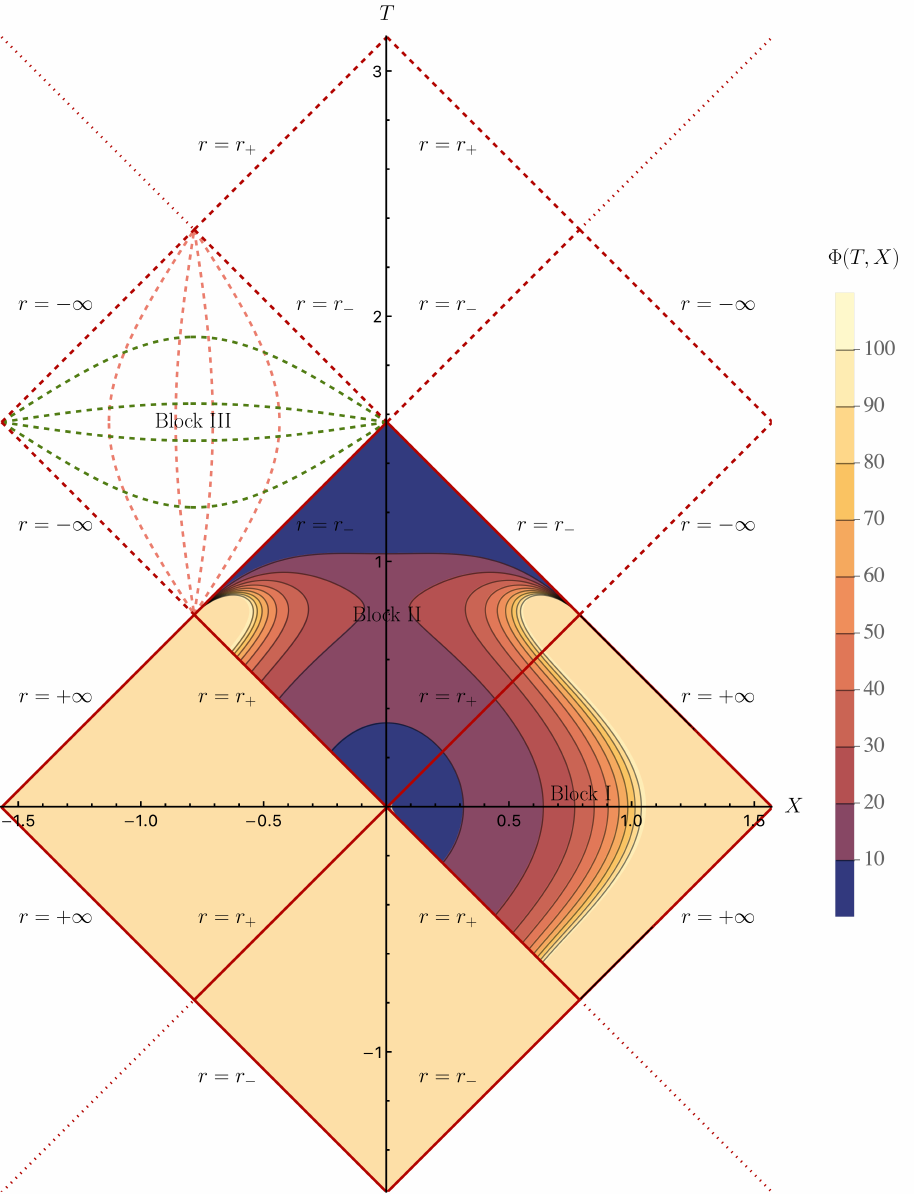}
\includegraphics[width=0.49\textwidth]{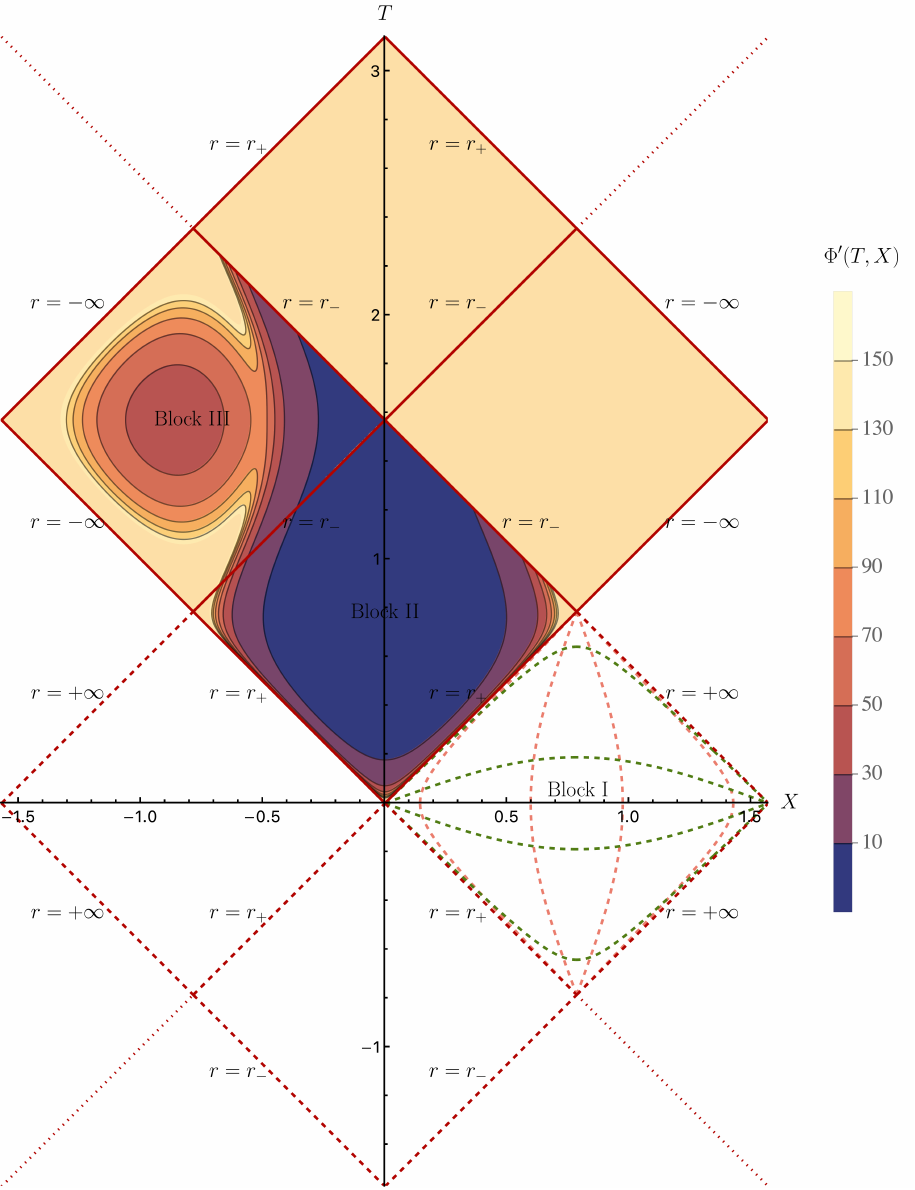}
\end{center}
\caption{\label{fig:conformal} Isolines of conformal factors $\Phi$ (left) and $\Phi'$ (right), corresponding to coordinate systems $K$ and $K'$, plotted in conformal diagrams of Fig.\ \ref{fig:penrose}.}
\end{figure}

Both diagrams shown in Fig.\ \ref{fig:penrose} are computed by a nearly standard Kruskal procedure (see, e.g., \cite{BoyerLindquist1967,Chrusciel2012}), but a few details should be given in order to explain the plots of hypersurfaces of constant time $t^\prime$.

A conformal diagram at the symmetry axis is constructed for the two dimensional metric
\begin{equation}
    {}^{(2)}g = - \left. \frac{\Delta}{\rho^2} \right|_{\cos^2 \theta = 1} dt^2 + \left. \frac{\rho^2}{\Delta} \right|_{\cos^2 \theta = 1} dr^2 = - F(r) dt^2 + \frac{1}{F(r)} dr^2,
\end{equation}
where
\begin{equation}
    F(r) := \frac{r^2 - 2 M r + a^2}{r^2 + a^2}.
\end{equation}
The usual Kruskal construction starts by defining a new coordinate
\begin{eqnarray}
r_\ast & := & \int^r \frac{dr^\prime}{F(r^\prime)} = \int^r \frac{{r^\prime}^2 + a^2}{{r^\prime}^2 - 2 M r^\prime + a^2} dr^\prime \nonumber \\
& = & r + \frac{M}{\sqrt{M^2 - a^2}} \left( r_+ \log \left| \frac{r - r_+}{M} \right| - r_- \log \left| \frac{r - r_-}{M} \right|\right).
\label{xiast}
\end{eqnarray}
Expression (\ref{xiast}) is well defined on $r \in \mathbb R$, except at $r = r_\pm$, where it diverges. Also note that by writing Eq.\ (\ref{xiast}) we explicitly set an integration constant, which in principle can be chosen separately in each of the Blocks I, II, and III.

A standard construction assumes the following coordinate transformations in each of the blocks:
\begin{equation}
    u := t - r_\ast, \quad v := t + r_\ast
\end{equation}
and
\begin{equation}
    \hat u := - \exp (- c u), \quad \hat v := \exp (c v),
\end{equation}
where $c$ is a constant.
This gives
\begin{equation}
    {}^{(2)}g = -F dt^2 + \frac{1}{F} dr^2 = -\frac{F}{c^2 \exp (2 c r_\ast)} d \hat u d \hat v,
\end{equation}
where in our case
\begin{equation}
    \exp(2 c r_\ast) = \exp(2 c r) \left| \frac{r - r_+}{M} \right|^\frac{2 c M r_+}{\sqrt{M^2 - a^2}} \left| \frac{r - r_-}{M} \right|^{-\frac{2 c M r_-}{\sqrt{M^2 - a^2}}}.
\end{equation}
The constant $c$ can be chosen in a way that allows for a regularization of the conformal factor $F/[c^2 \exp(2 c r_\ast)]$ at one of the horizons. This can be done by selecting $c = F^\prime(r_+)/2$ or $c = F^\prime(r_-)/2$. Indeed, one has
\begin{equation}
    \frac{F'(r_\pm)}{2} = \pm \frac{\sqrt{M^2 - a^2}}{r_\pm^2 + a^2} = \pm \frac{\sqrt{M^2 - a^2}}{2 M r_\pm}.
\end{equation}
In particular, selecting $c = F'(r_+)/2$, we get
\begin{equation}
    \exp(2cr_\ast) = \exp(2 c r) \left| \frac{r - r_+}{M} \right| \left| \frac{r - r_-}{M} \right|^{-r_-/r_+}.
\end{equation}
This choice will be used in the construction of a particular coordinate system, regular across Blocks I and II, and denoted further by $K$. It follows that
\begin{equation}
\label{conformalblock2}
    \frac{F}{c^2 \exp(2 c r_\ast)} = \pm \frac{M^2}{c^2 \exp(2 c r) (r^2 + a^2)} \left( \frac{r - r_-}{M} \right)^\frac{2 M}{r_+},
\end{equation}
where the signs $+$ and $-$ correspond to Blocks I and II, respectively. Note that expression (\ref{conformalblock2}) vanishes at the Cauchy horizon $r = r_-$.

Variables $\hat u$ and $\hat v$ can be compactified by setting
\begin{equation}
    U := \arctan(\hat u), \quad V := \arctan(\hat v).
\end{equation}
Finally, one defines Cartesian type coordinates $T$ and $X$, but this definition differs from one block to another. In Block I we set
\begin{equation}
    U = T - X, \quad V = T + X,
\end{equation}
or, equivalently,
\begin{equation}
    T := \frac{U + V}{2}, \quad X := \frac{V - U}{2}.
\end{equation}
so that
\begin{equation}
    - d \hat u d \hat v = \frac{1}{\cos^2(T - X) \cos^2(T + X)}(-dT^2 + dX^2).
\end{equation}
This yields the metric in the form
\begin{equation}
    {}^{(2)}g = -F dt^2 + \frac{1}{F} dr^2 = \Phi(T,X) (-dT^2 + dX^2),
\label{metricTX}
\end{equation}
where the conformal factor $\Phi$ reads
\begin{eqnarray}
    \Phi(T,X) & = & \frac{F}{c^2 \exp (2 c r_\ast) \cos^2(T - X) \cos^2(T + X)} \nonumber \\
    & = & \frac{M^2}{c^2 \exp(2 c r) (r^2 + a^2)} \left( \frac{r - r_-}{M} \right)^\frac{2 M}{r_+} \frac{1}{\cos^2(T - X) \cos^2(T + X)}.
\end{eqnarray}

In Block II one defines
\begin{equation}
    U = X - T, \quad V = T + X,
\end{equation}
and thus
\begin{eqnarray}
    - d \hat u d \hat v = - \frac{1}{\cos^2(T - X) \cos^2(T + X)}(-dT^2 + dX^2).
\end{eqnarray}
Consequently, we get in Block II,
\begin{equation}
  {}^{(2)}g  =  \frac{M^2}{c^2 \exp(2 c r) (r^2 + a^2)} \left( \frac{r - r_-}{M} \right)^\frac{2 M}{r_+} \frac{1}{\cos^2(T - X) \cos^2(T + X)} (-dT^2 + dX^2), 
\end{equation}
which coincides with expression (\ref{metricTX}). Thus, coordinates $(T,X)$ cover smoothly Blocks I and II; moreover, expression (\ref{metricTX}) remains regular at $r = r_+$. Unfortunately, coordinates $(T,X)$ defined above cannot be extended to Block III. At $r = r_-$ the conformal factor $F/[c^2 \exp(2 c r_\ast)]$ has a zero. On the other hand, the surface $r = r_-$ corresponds to $T - X = \pi/2$, and consequently the term $\cos^2(T - X)$ in the denominator of (\ref{metricTX}) vanishes at $r = r_-$. A careful inspection shows that the ratio of these two factors vanishes as well. This behaviour is shown in Fig.\ \ref{fig:conformal} (left panel), where we plot the conformal factor in Blocks I and II. It is also well known that the coordinates $(T,X)$ defined above can be extended to cover four different blocks, marked in yellow in Figs.\ \ref{fig:penrose} or \ref{fig:conformal} (left panels).

Joining Blocks II and III is possible by constructing a separate coordinate system, which we denote as $K'$, and which can also be extended to span across four different blocks (marked in yellow in right panels of Figs.\ \ref{fig:penrose} or \ref{fig:conformal}). It can be defined in an analogous way, but now instead of $c = F'(r_+)/2$, we take $c = F'(r_-)/2 < 0$. In this case, we get
\begin{equation}
\exp(2 c r_\ast) = \exp(2cr) \left| \frac{r - r_+}{M} \right|^{- \frac{r_+}{r_-}} \left| \frac{r - r_-}{M} \right|.
\end{equation}
Thus,
\begin{equation}
    \frac{F}{c^2 \exp(2 c r_\ast)} = \mp \frac{M^2}{c^2 \exp(2 c r) (r^2 + a^2)} \left( \frac{r_+ - r}{M} \right)^\frac{2M}{r_-},
\end{equation}
where the signs $-$ and $+$ correspond to Blocks II and III, respectively.

Coordinates $(T,X)$ are now defined in a way analogous to system $K$, but we shift $T$ by $\pi/2$. In Block II we define
\begin{equation}
    U = X - T + \frac{\pi}{2}, \quad V = T + X - \frac{\pi}{2},
\end{equation}
so that
\begin{equation}
    -d \hat u d \hat v = - \frac{1}{\cos^2(X - T + \pi/2) \cos^2(T + X - \pi/2)}(-dT^2 + dX^2).
\end{equation}
In Block III we set
\begin{equation}
    U = T - X - \frac{\pi}{2}, \quad V = T + X - \frac{\pi}{2},
\end{equation}
and
\begin{equation}
    -d \hat u d \hat v = \frac{1}{\cos^2(X - T + \pi/2) \cos^2(T + X - \pi/2)}(-dT^2 + dX^2).
\end{equation}
Combining the above results we get, in Blocks II and III, ${}^{(2)}g  = \Phi'(T,X) (-dT^2 + dX^2)$, where
\begin{equation}
\Phi'(T,X)  =  \frac{M^2}{c^2 \exp(2 c r) (r^2 + a^2)} \left( \frac{r_+ - r}{M} \right)^\frac{2 M}{r_-} \frac{1}{\cos^2(X - T + \pi/2) \cos^2(T + X - \pi/2)}.
\end{equation}
The conformal factor $\Phi'$ is plotted in Fig.\ \ref{fig:conformal} (right panel). In contrast to the coordinate patch $K$, the metric ${}^{(2)}g$ is now regular at $r = r_-$, but not at $r = r_+$.

Plotting the surfaces of constant $t$ and $r_\ast$ is possible in each of the blocks and in both coordinate systems with the help of following formulas. In patch $K$, we have
\begin{equation}
\label{constrastt}
    - \tanh \left( c r_\ast \right) = \frac{\cos (2 X)}{\cos (2 T)}, \quad \tanh (c t) = \frac{\sin (2 T)}{\sin (2 X)}
\end{equation}
in Block I and
\begin{equation}
    - \tanh (c r_\ast) = \frac{\cos (2 T)}{\cos (2 X)}, \quad \tanh (c t) = \frac{\sin (2 X)}{\sin (2 T)}
\end{equation}
in Block II. In patch $K'$ one gets
\begin{equation}
    \tanh (c r_\ast) = \frac{\cos (2 T)}{\cos (2 X)}, \quad - \tanh (c t) = \frac{\sin (2 X)}{\sin (2 T)}
\end{equation}
in Block II and
\begin{equation}
    \tanh \left( c r_\ast \right) = \frac{\cos (2 X)}{\cos (2 T)}, \quad - \tanh (c t) = \frac{\sin (2 T)}{\sin (2 X)}
\end{equation}
in Block III. Plotting the lines of constant $t^\prime$ is more involved. From Eq.\ (\ref{eqn:t.phi}), we have $t^\prime = t - r + r_\ast$. We plot the lines $t^\prime = \mathrm{const}$ by inverting, numerically, the relation $r_\ast = r_\ast(r)$, defined by Eq.\ (\ref{xiast}).

\end{document}